\documentclass[onefignum,onetabnum]{siamonline171218}

\usepackage{amsmath,amssymb,amsfonts}
\usepackage{mathtools}
\usepackage{subdepth} %
\usepackage{setspace}
\usepackage{booktabs,multirow}
\usepackage{graphicx}
\usepackage{epstopdf}
\usepackage[caption=false]{subfig}
\usepackage{xcolor}

\usepackage[ruled,vlined,algo2e]{algorithm2e}

\SetCommentSty{mycommfont}

\usepackage{microtype}
\usepackage{comment}

\usepackage{float}
\usepackage{xargs}                      %

\usepackage[colorinlistoftodos,prependcaption,textsize=tiny]{todonotes}

\usepackage[letterpaper,centering]{geometry}
\newcommand{\TheTitle}{A layered multiple importance sampling scheme for focused optimal Bayesian experimental design}
\newcommand{\TheAuthors}{C.~Feng and Y.~M.~Marzouk}

\headers{}{\TheAuthors}

\title{{\TheTitle}\thanks{Support from the AFOSR Computational
    Mathematics Program and by an AFOSR MURI, F.\ Fahroo, program manager.}}

\author{Chi Feng$^{*}$ \and Youssef M.\ Marzouk\thanks{Massachusetts Institute of Technology, Cambridge, MA (\email{chifeng@mit.edu}, \email{ymarz@mit.edu}; \url{http://uqgroup.mit.edu}).}}

\DeclareMathOperator{\Ex}{\mathbb{E}}

\DeclareMathOperator{\Var}{\mathbb{V}}
\DeclareMathOperator{\DKL}{D_{KL}}

\relax
\newcommand{\transpose}{{\top}}
\renewcommand{\d}{\text{d}}
\newcommand{\given}{\hspace{0.2ex}|\hspace{0.2ex}}
\newcommand{\RR}{\mathbb{R}}

\newcommand{\N}{\mathcal{N}}
\newcommand{\sigeps}{\sigma_\epsilon}

\newcommand{\vTheta}{{\boldsymbol{\Theta}}}
\newcommand{\vtheta}{\boldsymbol{\theta}}
\newcommand{\vGamma}{\boldsymbol{\Gamma}}

\newcommand{\vz}{\boldsymbol{z}}

\newcommand{\vy}{\boldsymbol{y}}
\newcommand{\vd}{\boldsymbol{d}}
\newcommand{\vG}{\boldsymbol{G}}
\newcommand{\veta}{\boldsymbol{\eta}}
\newcommand{\vEta}{\mathbf{H}}
\newcommand{\vepsilon}{\boldsymbol{\epsilon}}
\newcommand{\vY}{\boldsymbol{Y}}
\newcommand{\vmu}{\boldsymbol{\mu}}
\newcommand{\vSigma}{{\boldsymbol{\Sigma}}}

\newcommand{\SA}{{S_A}}
\newcommand{\SB}{{S_B}}
 
\newcommandx{\unsure}[2][1=]{\todo[linecolor=red,backgroundcolor=red!25,bordercolor=red,#1]{#2}}
\newcommandx{\change}[2][1=]{\todo[linecolor=blue,backgroundcolor=blue!25,bordercolor=blue,#1]{#2}}
\newcommandx{\info}[2][1=]{\todo[linecolor=OliveGreen,backgroundcolor=OliveGreen!25,bordercolor=OliveGreen,#1]{#2}}
\newcommandx{\improvement}[2][1=]{\todo[linecolor=Plum,backgroundcolor=Plum!25,bordercolor=Plum,#1]{#2}}

\hypersetup{
  pdftitle={\TheTitle},
  pdfauthor={\TheAuthors}
}

\begin{document}

\maketitle

\begin{abstract}
  We develop a new computational approach for ``focused'' optimal
  Bayesian experimental design with nonlinear models, with the goal of
  maximizing expected information gain in targeted \emph{subsets} of model
  parameters. Our approach considers uncertainty in the full set of
  model parameters, but employs a design objective that can exploit
  learning trade-offs among different parameter subsets.
  We introduce a new layered multiple importance sampling scheme
  that provides consistent estimates of expected information gain in
  this focused setting. This sampling scheme yields significant
  reductions in estimator bias and variance for a given computational
  effort, making optimal design more tractable for a wide range of
  computationally intensive problems.

\end{abstract}

\begin{keywords}
optimal experimental design, Bayesian inference, Monte Carlo methods,
multiple importance sampling, expected information gain, mutual
information
\end{keywords}

\begin{AMS}

\end{AMS}

\section{Introduction}
Data acquired from experiments are essential for developing and refining models. Acquiring data can be a costly process, so it is important to carefully choose experimental conditions that maximize the value of the acquired data, as defined by some \emph{optimal experimental design} criterion~\cite{atkinson:1992:oed}. In general, the design criterion should depend on what one intends to do with the data acquired from an experiment.

In this paper, we consider the optimal design of experiments where the observed data are used to infer uncertain model parameters. This setting includes models that may contain a large number of parameters, but where the user may care only about inference of a targeted subset. For example, in the context of chemical kinetics, experimental outcomes might be influenced by multiple rate constants and thermodynamic parameters, but only the rates of certain subsets of reactions may be of interest. Another example is the Kennedy \& O'Hagan model calibration framework~\cite{kennedy:2001:bco}, where users may favor experimental designs that focus on reducing uncertainties in `physical' parameters over reducing uncertainty in `tuning' parameters. To this end, we employ a design criterion that favors experiments maximizing the expected information gain in user-selected parameters of interest; following~\cite{DanLevineMITPhDthesis}, we refer to this approach as \emph{focused} optimal experimental design. Our key contribution is a new multiple importance sampling framework that makes focused experimental design tractable.

We formulate our experimental design problems in a Bayesian setting, where uncertain parameters are endowed with probability distributions and parameter inference can naturally be performed when the data are noisy and limited in number~\cite{bernardosmith2008,sivia2006}. We adopt a decision-theoretic approach, following \cite{lindley:1972:bsa}, wherein an optimal design maximizes the \emph{expected utility} of an experiment. As suggested by~\cite{lindley:1956:oam}, our utility function is chosen to be the information gain due to an experiment, which is equivalent to the Kullback--Leibler divergence from the prior to the posterior---or from selected prior marginals to the corresponding posterior marginals. This nonlinear Bayesian optimal design framework is a generalization of Bayesian 
D-optimal design, making no assumptions on the linearity of the forward model or the Gaussianity of the prior and noise distributions~\cite{chaloner:1995:bed}.

For models where the observables depend nonlinearly on the parameters of interest, exact evaluation of this design criterion is in general intractable, as it requires integration over the possible outcomes of the experiment and the space of model parameters. A nested Monte Carlo estimator \cite{rainforth2016nestedMC} of the expected information gain (jointly over \emph{all} model parameters) is proposed in~\cite{ryan:2003:eei}, but has significant limitations due to computational expense. Alternatives to this nested Monte Carlo estimator include kernel estimators of mutual information %
as proposed in \cite{kraskov2004estimating} and applied in \cite{terejanu2012bayesian}, but these approaches involve density estimation and thus require careful parameter choices to control bias and variance, along with large sample populations.
Laplace approximations to the posterior distribution~\cite{long2015123,bisetti2016optimal} or computationally efficient surrogate models~\cite{huan2013} can result in more tractable design criteria, but are unsuitable when these approximations break down, e.g., as the posterior distribution departs from normality or when a sufficiently accurate surrogate cannot be constructed. In general, these approximations induce asymptotic bias and do not yield consistent estimators of the expected information gain. Of course, it is possible to use Laplace approximations to construct consistent importance sampling estimators, as recently proposed in \cite{Beck2017} and discussed below; these techniques are complementary to the Monte Carlo methods developed here. Further reviews of modern computational methods and approximations for decision-theoretic Bayesian design can be found in \cite{RyanDrovandiMcGreePettitt_ISR2016,woods2017bayesian}. 

Other recent experimental design formulations avoid computing the expected information gain criterion altogether, and instead begin with the Bayesian A-optimality criterion~\cite{alexanderian2014optimal} for linear problems, extending it to nonlinear settings by (i) using Laplace approximations of the posterior for tractability; and (ii) averaging over many possible realizations of the data~\cite{alexanderian2016Aoptimal}. These approaches are similar to nonlinear design criteria that average functionals of the expected or observed Fisher information matrix, as described in~\cite{chaloner:1995:bed}. 
Though these alternatives provide computational scalability and can be rationalized in weakly nonlinear problems, they remain only approximations of a decision-theoretic expected utility. Also, while D-optimal design and its generalization to expected information gain are invariant under reparameterization of the underlying model, A-optimal design and its extensions discussed above are not~\cite{biedermann2011optimal,dette1997designing}.

This paper proposes a consistent, yet tractable Monte Carlo estimator of the expected information gain, not limited by linear 
 approximations. Estimating the expected information gain requires estimating posterior normalizing constants for many realizations of the data. Yet estimating the posterior normalizing constant---even for a single realization of the data---is a challenging task. In addition, our criterion for \emph{focused} experimental design introduces marginalization over nuisance parameters~\cite{Stumpf2013marginal}, further reducing the efficiency and accuracy of existing Monte Carlo approaches. While many schemes~\cite{chib2001marginal,gelman1998} have been developed to estimate posterior normalizing constants, they only target a single posterior distribution. Estimating posterior normalizing constants over many realizations of the data, however, allows for the sharing of information between estimators. We use this idea to develop a \emph{layered multiple importance sampling} (LMIS) estimator of the expected information gain, which iteratively improves the biasing distributions used to estimate posterior normalizing constants. This is accomplished by optimally combining samples and cached model evaluations from earlier iterations using multiple importance sampling~\cite{veach1995}, yielding consistent and more accurate estimators.

Our approach has roots in the simpler importance sampling scheme of~\cite{FengSMthesis}. A useful generalization of the latter, recently proposed by~\cite{Beck2017,Englezou2018}, involves computing a Laplace approximation of the posterior for each realization of the data, and using this approximation to construct a biasing distribution to estimate each posterior normalizing constant. The resulting expected information gain estimators are also consistent, but require solving an optimization problem and computing a Hessian for each realization. In contrast, the LMIS scheme developed here emphasizes \emph{recycling} information from one realization of the data to the next. Within this framework, many different biasing schemes can be used. The particular biasing schemes employed here avoid repeated optimization and do not require gradients or Hessians of the forward model or log-likelihood. LMIS is thus entirely non-intrusive, in the sense of requiring only pointwise evaluations of the likelihood function.

We demonstrate using both linear and nonlinear numerical examples examples that our LMIS estimator offers multiple order-of-magnitude reductions in mean squared error over comparable Monte Carlo methods using the same number of model evaluations, while remaining asymptotically exact. This new estimator enables efficient estimation of expected information gain for both joint and the more challenging case of focused experimental design.

This paper is organized as follows. In \cref{sec:oed}, we introduce the Bayesian optimal experimental design problem and formulate the focused information gain criterion. We illustrate this formulation using a linear-Gaussian toy problem---which has closed-form expressions of the expected information gain, and a clear trade-off between inferring all the model parameters and inferring a subset thereof. In \cref{sec:mc} we derive a nested Monte Carlo estimator of our focused design criterion, allowing for a generic choice of importance sampling scheme in its inner loops; we then analyze the asymptotic bias and variance of this estimator. \Cref{ssec:lmis} describes our layered multiple importance
 sampling scheme. \Cref{sec:numerical-results} evaluates the numerical performance of the LMIS estimator on linear problems of varying dimension, and in \cref{ssec:mossbauer} we apply the LMIS algorithm to the optimal design of a M\"ossbauer spectroscopy experiment, which is nonlinear in its parameters.

\section{Optimal experimental design}%
\label{sec:oed}
The Bayesian paradigm treats unknown parameters as random variables. Let $(\Omega,\mathcal{F},\mathbb{P})$ be a probability space, where $\Omega$ is a sample space, $\mathcal{F}$ is a $\sigma$-field, and $\mathbb{P}$ is a probability measure on $(\Omega,\mathcal{F})$. Let the real-valued random vector $\vz \,:\,\Omega\to\RR^{n_\theta + n_\eta}$ represent the uncertain parameters of a statistical model, and let it be partitioned coordinate-wise, $\vz \coloneqq (\vtheta, \veta)$, into $\vtheta\,:\,\Omega\to\RR^{n_\theta}$ and $\veta\,:\,\Omega\to\RR^{n_\eta}$, which denote the \emph{parameters of interest} and \emph{nuisance parameters}, respectively. Both $\vtheta$ and $\veta$ are parameters to be conditioned on experimental data.
The random variable $\vz$ is associated with a prior measure $\mu$ on $\mathbb{R}^{n_\theta+n_\eta}$ such that $\mu(A)=\mathbb{P}(\vz^{-1}(A))$ for $A\in\mathbb{R}^{n_\theta + n_\eta}$. We can then define $p(\vz) = \text{d}\mu/\text{d}\vz = p(\vtheta, \veta) $ as the joint density of $\vtheta$ and $\veta$ with respect to the Lebesgue measure $\text{d}\vz$. For the present purposes, we will assume that this density always exists. Similarly, the data $\vy$ can be treated a real-valued random variable $\vy\,:\,\Omega\to\RR^{n_y}$. Finally, let $\vd\in \mathcal{D} \subseteq \RR^{n_d}$ denote the \emph{design variables}, or experimental conditions. Hence, $n_\theta$ is the number of uncertain parameters of interest, $n_\eta$ is the number of nuisance parameters, and $n_d$ is the number of design variables. In the Bayesian setting, upon observing a realization of the data $\vy$ obtained by performing an experiment under conditions $\vd$, we can update our knowledge about the model parameters using Bayes' rule:
\begin{equation}
\nonumber
\underbrace{p(\vtheta, \veta \vert \vy,\vd)}_{\text{posterior}}=\underbrace{p(\vy | \vtheta, \veta, \vd)}_{\text{likelihood}}\,\underbrace{p(\vtheta, \veta)}_{\text{prior}}/\underbrace{p(\vy| \vd)}_\text{evidence}.
\label{eq:bayes-rule}
\end{equation}
Here $p(\vtheta,\veta )$ is the prior density, $p(\vy| \vtheta, \veta, \vd)$ is the likelihood function associated with our statistical model, $p(\vtheta, \veta | \vy, \vd)$ is the posterior density, and $p(\vy| \vd)$ is the evidence or posterior normalizing constant. It is reasonable to assume that prior knowledge on $\vtheta$ and $\veta$ (i.e., the prior measure $\mu$) does not depend on the experimental design, and thus we employ the simplification $p(\vtheta,\veta| \vd)=p(\vtheta,\veta)$.

The likelihood function often combines a deterministic \emph{forward model} with a statistical model for measurement and/or model errors. For example, the observations might be modeled as additive Gaussian perturbations of some deterministic model predictions $\vG(\vtheta, \veta, \vd)$:
\begin{equation*}
\vy = \vG(\vtheta, \veta, \vd) + \vepsilon, \quad \vepsilon \sim \mathcal{N}(0, \Gamma_\epsilon). 
\end{equation*}
Here, $\vG$ is the forward model; we will assume that it is (in general) nonlinear and that its evaluations are the dominant computational cost of evaluating or simulating from the likelihood. Thus, while evaluations of $\vG(\vtheta, \veta, \vd)$ for new values of the parameters or design variables are expensive, it is cheap to evaluate $p(\vy \vert \vtheta, \veta, \vd)$ for different values of $\vy$, given fixed values of $\vtheta$, $\veta$, and $\vd$. This assumption is representative of many science and engineering problems, where $\vG$ might represent the solution of a set of differential equations and where the noise model is relatively simpler.

\subsection{\label{sec:expected-utility}Expected utility framework}
Following the decision-theoretic approach of Lindley \cite{lindley:1972:bsa}, an objective for optimal experimental design should have the following general form:
\begin{align}
U(\vd) &= \int_{\mathcal{Y}}\int_{\vTheta}\int_{\vEta}u(\vy,\vd,\vtheta,\veta)\,p(\vtheta,\veta,\vy|\vd)\,\d{\vtheta}\,\d{\veta}\,\d{\vy},
\label{eq:expected-utility}
\end{align}
where $u(\vy,\vd,\vtheta,\veta)$ is a \emph{utility function} representing the value of observing a particular outcome $\vy$ of the experiment under design $\vd$, when the parameters are $\vtheta$ and $\veta$. We do not know the outcome of the experiment \textit{a priori}, however; nor do we know the true values of the parameters $\vtheta$ and $\veta$. So we instead consider the \emph{expected} utility $U(\vd)$, where the expectation is taken over the joint distribution of $\vtheta$, $\veta$, and $\vy$, and where $\vTheta \subseteq \mathbb{R}^{n_\theta}$, $\vEta \subseteq \mathbb{R}^{n_\eta}$ and $\mathcal{Y} \subseteq \mathbb{R}^{n_y}$ are the ranges of $\vtheta$, $\veta$, and $\vy$, respectively.

To \emph{focus} attention on inference of the parameters $\vtheta$, we choose $u(\vy,\vd,\vtheta,\veta)$ to be the Kullback--Leibler (KL) divergence (i.e., relative entropy) from the \emph{marginal} prior on $\vtheta$ to the \emph{marginal} posterior of $\vtheta$, where the marginalization is over the nuisance parameters $\veta$. In other words, the divergence is from the marginal prior on the parameters of interest,
\begin{align*}
p_\theta(\vtheta) = \int_{\vEta}p(\vtheta,\veta)\,\text{d}\veta,
\end{align*}
to the marginal posterior on the parameters of interest,
\begin{align*}
p_\theta(\vtheta|\vy,\vd) =
  \int_{\vEta}p(\vtheta,\veta| \vy,\vd)\text{d}{\veta},
  \label{eq:marginal-posterior}
\end{align*}
and hence our utility function is
\begin{align}
\nonumber u(\vy,\vd,\vtheta,\veta) &\coloneqq \DKL\left(p_\theta(\vtheta| \vy,\vd)\,\bigr\|\, p_\theta(\vtheta)\right) \\
&= \int_{\vTheta} p_\theta(\tilde{\vtheta}| \vy,\vd) \log \biggl[\frac{p_\theta(\tilde{\vtheta}| \vy,\vd)}{p_\theta(\tilde{\vtheta})}\biggr]\text{d}{\tilde{\vtheta}}  = u(\vy, \vd).  \label{eq:information-metric} 
\end{align}
Above, $\tilde{\vtheta}$ is simply a dummy variable of integration. The final equality reflects the fact that the KL divergence does not actually depend on the values of $\vtheta$ and $\veta$. Thus we can write the expected utility (\ref{eq:expected-utility}) of a design as an expectation over the data only:
\begin{align}
  \nonumber
  U(\vd) &=  \int_{\vY}\int_{\vTheta}\int_{\vEta}u(\vy,\vd)\,p(\vtheta,\veta,\vy|\vd)\text{d}\vtheta\text{d}\veta\text{d}\vy\\
  &=  \int_{\vY} u(\vy,\vd) p(\vy| \vd) \text{d}{\vy}.
  \label{eq:objective-function}
\end{align}
Now, substituting the utility (\ref{eq:information-metric}) into the objective function (\ref{eq:objective-function}), we obtain
\begin{align}
  \nonumber
  U(\vd)&= \int_{\vY}\int_{\vTheta} p_\theta(\vtheta| \vy,\vd) \log \biggl[\frac{p_\theta(\vtheta| \vy,\vd)}{p_\theta(\vtheta)}\biggr]\text{d}\vtheta\,p(\vy| \vd)\text{d}{\vy} \nonumber\\
  &= \Ex_{\vy| \vd}\Bigl[\DKL\bigl(p_\theta(\vtheta| \vy,\vd)\,\bigr\|\,p_\theta(\vtheta)\bigr)\Bigr],
  \label{eq:objective-function-simple}
\end{align}
which means that the expected utility $U(\vd)$ is the \emph{expected information gain} in the parameters of interest $\vtheta$, irrespective of the information gain in the nuisance parameters $\veta$. Intuitively, the expected information gain describes the average reduction in entropy from the prior marginal of $\vtheta$ to the posterior marginal of $\vtheta$, where the average is taken over all possible experimental outcomes $\vy$, weighed according to their prior predictive distribution.  $U(\vd)$ is equivalent to the \emph{mutual information}
between the parameters of interest $\vtheta$ and the data $\vy$, given a particular design $\vd$ \cite{CoverThomas}.

\subsection{Linear--Gaussian example}
\label{ssec:linear-gaussian}
To illustrate our formulation for focused optimal experimental design, we first apply it to a toy problem that can be solved analytically. Consider the following experiment with a {linear} model,
\begin{equation*}
  \begin{bmatrix}y_1\\y_2\end{bmatrix}=\begin{bmatrix}d&0\\0&1-d\end{bmatrix}\begin{bmatrix}\theta\\\eta\end{bmatrix}+\begin{bmatrix}\epsilon_1\\\epsilon_2\end{bmatrix},
  \label{eq:lin-observation-model}
\end{equation*}
additive independent noise $\epsilon_1,\epsilon_2\sim\N(0,\sigma_\epsilon^2)$, and a design space $d\in[0,1]$. The prior distributions on $\theta$ and $\eta$ are chosen to be independent standard normals, $\theta\sim\N(0,1)$ and $\eta\sim\N(0,1)$. The posterior $\theta,\eta |\vy,d \sim \N(\vmu_\text{post}, \vSigma_\text{post})$ is a multivariate normal with mean and covariance available in closed form.
The marginal posterior $p( \theta  | \vy,d) $ is also Gaussian and follows immediately.

\begin{figure}[htbp]
    \centering
    \includegraphics[width=0.61\linewidth]{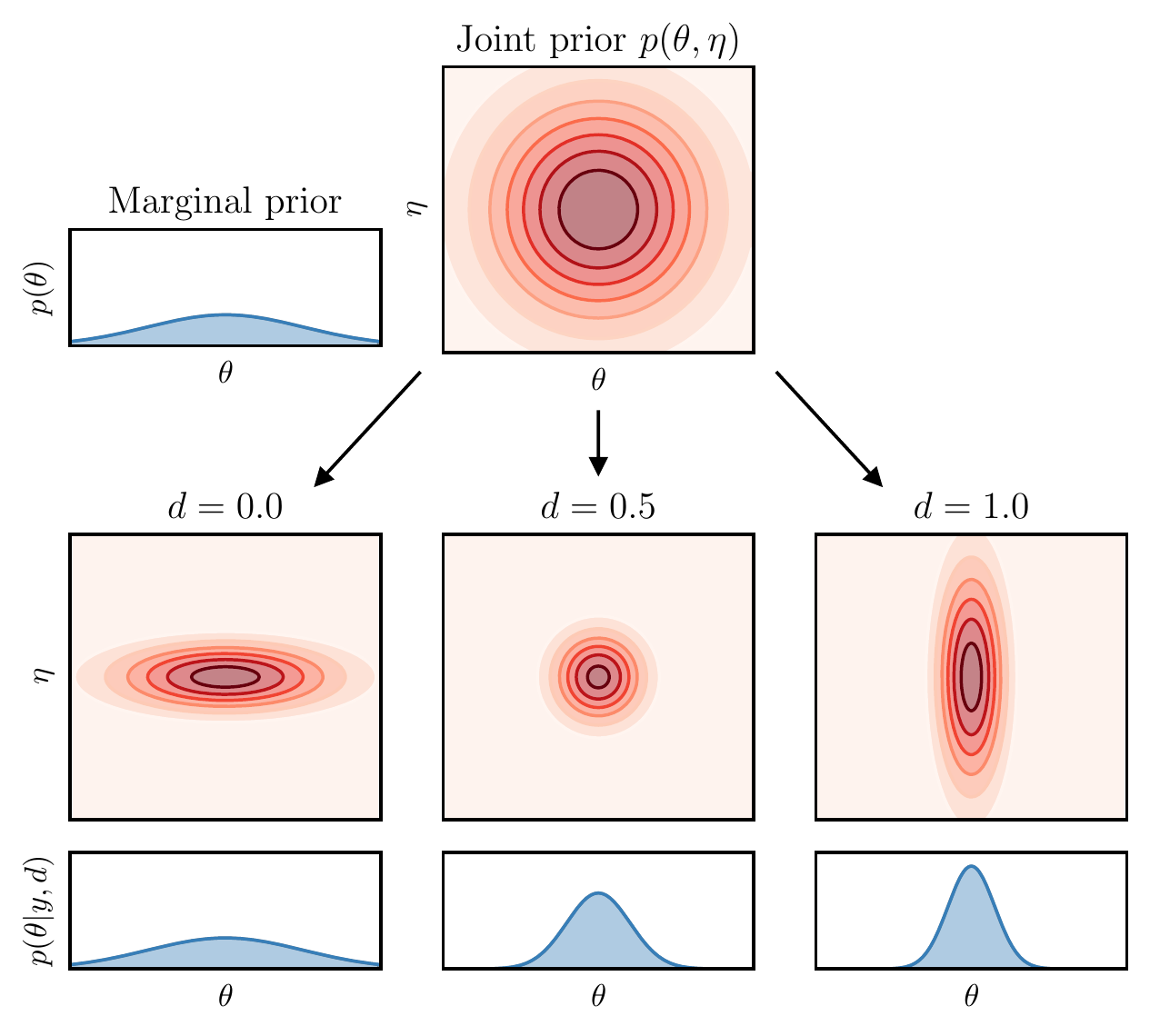}
    \quad
    \includegraphics[width=0.3\linewidth]{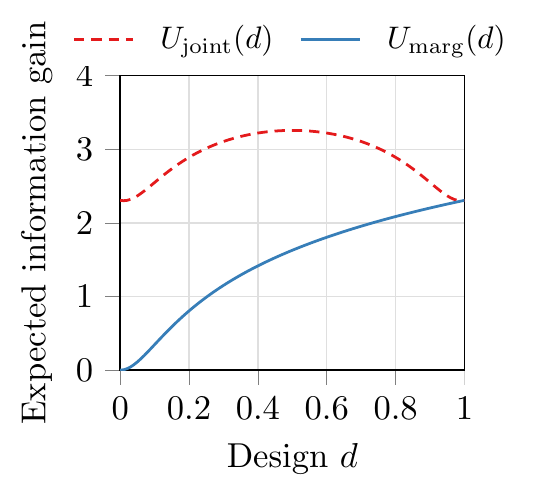}
    \caption{\emph{(Left)} Joint and marginal posteriors of the 2D linear Gaussian example. \emph{(Right)} Expected information gain for the parameter of interest alone (`marginal') and for all parameters (`joint').}
    \label{fig:linear-gaussian-eig}
\end{figure}

The structure of this model encodes a simple trade-off between learning about $\theta$ and learning about $\eta$. As $d$ increases, we obtain a larger signal-to-noise ratio in the measurement of $\theta$ while the signal-to-noise ratio associated with measuring $\eta$ decreases. This trade-off can be visualized by observing the anisotropy of the posterior distribution when varying $d$ for a fixed value of $\vy$, as illustrated in \cref{fig:linear-gaussian-eig}. The marginals at the bottom of the figure suggest that the design minimizing uncertainty in $\theta$ is simply $d=1$.

Since we are working with a linear-Gaussian model, we can obtain analytical expressions for the 
expected information gain in $(\theta, \eta)$ \emph{jointly},
\begin{align}
U_{\theta,\eta}(d)&=\Ex_{\vy|d}\bigl[\DKL\bigl(p (\theta,\eta | \vy,d ) \Vert p(\theta,\eta)\bigr)\bigr]=\frac{1}{2}\log\biggl[\frac{\bigl({(1-d)}^2+\sigeps^2\bigr)\bigl(d^2+\sigeps^2\bigr)}{\sigeps^4}\biggr],\label{eq:eig-joint}
\intertext{and in the parameter of interest $\theta$ alone,}
U_\theta(d)&=\Ex_{\vy|d}\bigl[\DKL\bigl(p(\theta|\vy,d) \Vert p(\theta)\bigr)\bigr]=\frac{1}{2}\log\biggl[1+\frac{d^2}{\sigeps^2}\biggr].\label{eq:eig-marg}
\end{align}
The expected information gains for both cases are plotted on the right of \cref{fig:linear-gaussian-eig}.

To find the optimal experimental design, the expected utility must be maximized over the design space $\mathcal{D} = [0,1]$. When the objective is to maximize information gain in the parameters of interest, we have $U = U_{\theta}$ and the optimal design is $d_\theta^\ast = 1$; this is consistent with the $\theta$ marginals shown in \cref{fig:linear-gaussian-eig}. On the other hand, when the objective is to maximize information gain in all uncertain parameters, we have $U = U_{\theta, \eta}$ and the optimal design becomes $d_{\theta,\eta}^\ast = \frac{1}{2}$. Notice that maximizing the expected information gain in \cref{eq:eig-joint} is equivalent to minimizing the determinant of the posterior covariance matrix, i.e., Bayesian D-optimal design \cite{chaloner:1995:bed}.
We will revisit more complex versions of this example in \cref{ssec:hd-linear-gaussian}, where we benchmark the convergence of various Monte Carlo estimators of \eqref{eq:objective-function-simple}.

\section{Monte Carlo estimation of the expected information gain}%
\label{sec:mc}
In general, the expected information gain (EIG)---whether in the parameters $\vtheta$ alone or jointly in $(\vtheta,\veta)$---does not have a closed form and must be approximated numerically, e.g., by a Monte Carlo estimate. In this section, we derive a nested Monte Carlo estimator of the focused EIG \eqref{eq:objective-function-simple}, analyze its properties, and then propose a layered multiple importance sampling (LMIS) estimator that has better properties. To make the subsequent developments clear, we note that the nested estimators employed here comprise an \emph{outer loop} that involves taking an expectation over the prior distribution and an \emph{inner loop} that involves computing certain posterior normalizing constants. \Cref{ssec:nestedMC} presents a general format for nested estimators of \eqref{eq:objective-function-simple} that employ importance sampling in their inner loops, but deliberately leaves the choice of biasing distributions open. \Cref{ssec:biasvariancescaling} analyzes the bias and variance of these estimators; this analysis also suggests what good biasing distributions ought to be. \Cref{ssec:scaling} discusses the convergence rate of the nested estimators and the optimal asymptotic scaling of inner loop and outer loop sample sizes.
\Cref{ssec:lmis} then presents a way of adaptively identifying biasing distributions via multiple importance sampling, which leads to our new LMIS estimator. LMIS can offer significant increases in performance over simpler and non-adaptive choices of biasing, while preserving asympotically exact estimation of the focused EIG.

\subsection{A nested importance sampling estimator}
\label{ssec:nestedMC}
The EIG as written in \eqref{eq:objective-function-simple} contains an inner loop of integration over the posterior distribution, and also requires evaluating the normalized posterior density of $\vtheta$; both are computationally undesirable. By applying Bayes' rule to the terms involving the posterior density, we can rewrite \eqref{eq:objective-function-simple} so that integration is instead performed over the prior distribution of $\vtheta$, $\veta$, and $\vy$:
\begin{align*}
    U(\vd) & = \int_{\vY} \left ( \int_{\vTheta}\log\biggl[\frac{p(\vtheta\given \vy,\vd)}{p(\vtheta)}\biggr]\,p(\vtheta\given \vy,\vd)\,\d{\vtheta} \right ) \,p(\vy\given \vd)\,\d{\vy} \\
           & =
    \int_{\vY}\int_{\vTheta} \int_{\vEta} \Bigl\{\log\bigl[p(\vy\given\vtheta,\vd)\bigr]-\log\bigl[p(\vy\given\vd)\bigr]\Bigr\}\,p(\vy\given\vtheta,\veta, \vd)\,p(\vtheta, \veta)\,\d{\veta}\,\d{\vtheta}\,\d{\vy}.
\end{align*}
This form of the focused EIG can be approximated by the following Monte Carlo estimate:
\begin{align}
    \widetilde{U}(\vd) = \frac{1}{N}\sum_{i=1}^N \log\bigl[\underbrace{p(\vy^{(i)}\given \vtheta^{(i)},\vd)}_{\substack{\text{conditional} \\\text{likelihood}}}\bigr]-\log\bigl[\underbrace{p(\vy^{(i)}\given \vd)}_{\substack{\text{marginal}\\\text{likelihood}}}\bigr],\label{eq:expected-utility-mc}
\end{align}
where the samples $\{\vtheta^{(i)}, \vy^{(i)} \}_{i=1}^N$ are drawn i.i.d.\ from the prior as follows: for each $i$, draw $(\vtheta^{(i)}, \veta^{(i)}) \sim p(\vtheta,\veta)$ from the joint prior on the parameters; then draw $\vy^{(i)} \sim p(\vy\given \vtheta^{(i)},\veta^{(i)},\vd)$. The sample $\veta^{(i)}$ is necessary to realize a value of $\vy^{(i)}$ but does not appear explicitly in \eqref{eq:expected-utility-mc}.

The summand in \eqref{eq:expected-utility-mc} contains the posterior normalizing constant $p(\vy^{(i)}\given \vd)$ for a given realization of the data $\vy^{(i)}$ and design parameters $\vd$, known as the \emph{marginal likelihood}, along with a partial posterior normalizing constant $p(\vy^{(i)}\given \vtheta^{(i)},\vd)$, which we call the \emph{conditional likelihood}. The marginal likelihood typically does not have an analytical form, but can be approximated by an importance sampling estimate for each $\vy^{(i)}$:
\begin{align}
    \nonumber
    p(\vy^{(i)}\given \vd) & =  \int_{\vTheta\times\vEta}p(\vy^{(i)}\given {\vtheta},{\veta},\vd)\,p({\vtheta},{\veta})\,\d{({\vtheta},{\veta})} \\
                           & \approx \frac{1}{M_1}\sum_{j=1}^{M_1} p(\vy^{(i)}\given \vtheta^{(i,j)},\veta^{(i,j)},\vd)\,w_\text{marg}^{(i,j)}
    \eqqcolon \widehat{p}(\vy^{(i)}\given\vd),
    \label{eq:marginal-likelihood-estimator}
\end{align}
where the samples $\{ \vtheta^{(i,j)}, \veta^{(i,j)}\}_{j=1}^{M_2}$ are drawn i.i.d.\ from a suitable biasing distribution $q_\text{marg}^{(i)}(\vtheta, \veta)$ %
with corresponding importance weights:
\begin{align}
    w_\text{marg}^{(i,j)} = \frac{p(\vtheta^{(i,j)},\veta^{(i,j)})}{q_\text{marg}^{(i)}(\vtheta^{(i,j)},\veta^{(i,j)})}.\label{eq:w_marg}
\end{align}
Similarly, the conditional likelihood evaluated at $\vy^{(i)}$, $p(\vy^{(i)}\given \vtheta^{(i)},\vd)$, can be estimated as follows:
\begin{align}
    \label{eq:condlike-true}
    p(\vy^{(i)}\given \vtheta^{(i)},\vd) & = \int_{\vEta} p(\vy^{(i)}\given \vtheta^{(i)},{\veta},\vd)\,p({\veta}\given \vtheta^{(i)})\,\d{{\veta}}        \\
                                         & \approx \frac{1}{M_2}\sum_{k=1}^{M_2} p(\vy^{(i)}\given \vtheta^{(i)},\veta^{(i,k)},\vd)\,w_\text{cond}^{(i,k)}
    \eqqcolon \widehat{p}(\vy^{(i)}\given\vtheta^{(i)},\vd),
    \label{eq:conditional-likelihood-estimator}
\end{align}
where samples $\{\veta^{(i,k)}\}_{k=1}^{M_2}$ are drawn i.i.d.\ from a biasing distribution $q_\text{cond}^{(i)}(\veta)$, with importance weights
\begin{align}
    w_\text{cond}^{(i,k)} = \frac{p(\veta^{(i,k)}\given \vtheta^{(i)})}{q_\text{cond}^{(i)}(\veta^{(i,k)})}.\label{eq:w_cond}
\end{align}
Substituting the above importance sampling estimates of the marginal and conditional likelihoods into \eqref{eq:expected-utility-mc}, we arrive at a nested Monte Carlo estimator for the EIG in $\vtheta$:
\begin{equation}
    \begin{multlined}
        \widehat{U}(\vd)=\frac{1}{N}\sum_{i=1}^N \biggl \{
        \log\Bigl[ \frac{1}{M_2}\sum_{k=1}^{M_2} p(\vy^{(i)}\given \vtheta^{(i)},\veta^{(i,k)},\vd)\,w_\text{cond}^{(i,k)} \Bigr] \\ \qquad
        -\log\Bigl[ \frac{1}{M_1}\sum_{j=1}^{M_1} p(\vy^{(i)}\given \vtheta^{(i,j)},\veta^{(i,j)},\vd)\,w_\text{marg}^{(i,j)}\Bigr] \biggr\}. \label{eq:big-mc-estimator}
    \end{multlined}
\end{equation}

So far, we have not explicitly specified the biasing distributions $\{ q_\text{marg}^{(i)} \}_{i=1}^N$ and $\{ q_\text{cond}^{(i)} \}_{i=1}^N$. In the next sections, we will identify optimal biasing distributions and describe methods to approximate them.

\subsection{Bias and variance of the nested estimator}%
\label{ssec:biasvariancescaling}

Choosing an appropriate biasing distribution requires understanding the bias and variance of the focused EIG estimator~\eqref{eq:big-mc-estimator}. Using the second-order delta method (see \cref{sec:appendix}), we find that the bias of $\widehat{U}(\vd)$ is, to leading order:
\begin{align}
    \Ex[\widehat{U}(\vd)-U(\vd)] & \approx \frac{1}{M_1}{\Ex_{\vy}\biggl[\frac{\sigma^2_\text{marg}(\vy,\vd)}{p^2(\vy|\vd)}\biggr]} - \frac{1}{M_2}{\Ex_{\vy,\vtheta}\biggl[\frac{\sigma^2_\text{cond}(\vy,\vtheta,\vd)}{p^2(\vy|\vtheta,\vd)}\biggr]}\label{eq:delta-bias} \\\nonumber
                                 & = \frac{C_1(\vd)}{M
        _1}-\frac{C_2(\vd)}{M_2},
\end{align}
and that its variance is, to leading order:
\begin{align}
    \Var[\widehat{U}(\vd)] & \approx \frac{1}{N}{\Var_{\vy,\vtheta}\biggl[\ln\frac{p(\vy|\vtheta,\vd)}{p(\vy|\vd)}\biggr]}+\frac{1}{NM_1}{\Ex_{\vy}\biggl[\frac{\sigma^4_\text{marg}(\vy,\vd)}{p^4(\vy|\vd)}\biggr]} + \frac{1}{NM_2}{\Ex_{\vy,\vtheta}\biggl[\frac{\sigma^4_\text{cond}(\vy,\vtheta,\vd)}{p^4(\vy|\vtheta,\vd)}\biggr]} \label{eq:delta-variance} \\\nonumber
                           & = \frac{D_3(\vd)}{N}+\frac{D_1(\vd)}{NM_1}+\frac{D_2(\vd)}{NM_2}.
\end{align}
The higher-order terms in the bias \eqref{eq:delta-bias} are $\mathcal{O}(1/M_1^2)$ and $\mathcal{O}(1/M_2^2)$, while the higher-order terms in the variance \eqref{eq:delta-variance} are $\mathcal{O}(1/(NM_1^2))$ and $\mathcal{O}(1/(NM_2^2))$; see \cref{sec:appendix}  for details.  
Also, $\sigma_\text{marg}^2(\vy,\vd)$ and $\sigma_\text{cond}^2(\vy,\vtheta,\vd)$ are variances of the summands in the marginal and conditional likelihood estimators \eqref{eq:marginal-likelihood-estimator} and \eqref{eq:conditional-likelihood-estimator}, respectively:
\begin{subequations}
    \begin{align}
        \sigma_\text{marg}^2(\vy,\vd)                                                                                                                                                                                            & \coloneqq
        \Var_{\widetilde{\vtheta},\widetilde{\veta}}\biggl[p(\vy|\widetilde{\vtheta},\widetilde{\veta},\vd)\frac{p(\widetilde{\vtheta},\widetilde{\veta})}{q_\text{marg}^{(\vy)}(\widetilde{\vtheta},\widetilde{\veta})}\biggr], & (\widetilde{\vtheta},\widetilde{\veta})\sim q_\text{marg}^{(\vy)},\label{eq:sigma_marg} \\
        \sigma_\text{cond}^2(\vy,\vtheta,\vd)                                                                                                                                                                                    & \coloneqq
        \Var_{\widetilde{\veta}}\biggl[p(\vy|\vtheta,\widetilde{\veta},\vd)\frac{p(\widetilde{\veta}|\vtheta)}{q_\text{cond}^{(\vy, \vtheta)}(\widetilde{\veta})}\biggr],                                                        & \widetilde{\veta}\sim q_\text{cond}^{(\vy, \vtheta)}.\label{eq:sigma_cond}
    \end{align}
\end{subequations}

From \eqref{eq:delta-bias}, we see that the bias is controlled by the variance of the inner marginal and conditional likelihood estimators, which can be reduced by choosing better biasing distributions $q_\text{marg}^{(\vy)}$ and $q_\text{cond}^{(\vy,\vtheta)}$ or by increasing the number of inner Monte Carlo samples $M_1$ and $M_2$.\footnote{
Here and in \eqref{eq:sigma_marg}--\eqref{eq:sigma_cond}, we write the inner biasing distributions as $q_\text{marg}^{(\vy)}$ and $q_\text{cond}^{(\vy, \vtheta)}$ to indicate that they can depend arbitrarily on the outer values of $\vy$ and possibly $\vtheta$. To recover the simpler notation of \cref{ssec:nestedMC}, where $i$ indexes the outer loop, simply put $q_\text{marg}^{(i)} =  q_\text{cond}^{(\vy^{(i)} )}$ and $q_\text{cond}^{(i)} =  q_\text{cond}^{ (\vy^{(i)}, \vtheta^{(i)} ) }$.}
The bias can take both positive and negative values since it is the difference of two positive quantities. This property contrasts with the nested Monte Carlo estimator of \emph{unfocused} EIG in \cite{ryan:2003:eei}, where the bias only takes positive values. The latter corresponds to the case of no nuisance parameters ($n_\eta = 0$); then the conditional likelihood $p(\vy|\vtheta,\vd)$ \eqref{eq:condlike-true} can be evaluated exactly, so $\sigma_\text{cond}^2(\vy, \vtheta, \vd)$ is effectively zero and \eqref{eq:delta-bias} is always positive.

The variance of $\widehat{U}(\vd)$ decreases with the number of outer samples as $1/N$. The first term, $D_3(\vd)$, depends only on the variability of the conditional and marginal likelihoods given $(\vy, \vtheta)$ drawn from the prior. The second and third terms, $D_1(\vd)$ and $D_2(\vd)$, depend explicitly on the variance of the inner estimators. Thus, reducing the variance of the inner estimators will also reduce the variance of $\widehat{U}(\vd)$.

In order to minimize the bias and variance of $\widehat{U}(\vd)$, we therefore need to find biasing distributions $q_\text{marg}^{(\vy)}$ and $q_\text{cond}^{(\vy,\vtheta)}$ that minimize the variance of the importance sampling estimates of the marginal and conditional likelihood. From \cref{eq:sigma_marg} and \cref{eq:sigma_cond}, we see that the using the joint and conditional posteriors as biasing distributions (with \emph{normalized} densities) would result in zero-variance estimators. Current approaches for evaluating the marginal likelihood in this setting, such as \cite{huan2013,ryan:2003:eei}, instead use the prior distribution $p(\vtheta,\veta)$ as a biasing distribution, i.e.,
\begin{align*}
    q_\text{marg}^{(\vy)}(\vtheta,\veta)=p(\vtheta,\veta)\quad\text{and}\quad
    q_\text{cond}^{(\vy,\vtheta)}(\veta)=p(\veta|\vtheta)\, ,
\end{align*}
so that the importance weights $w_\text{marg}$ and $w_\text{cond}$ in \eqref{eq:w_marg} and \eqref{eq:w_cond} are identically one. However, when the joint or conditional posterior distributions are more concentrated than their corresponding prior distributions, e.g., when the data $\vy^{(i)}$ are informative, using the prior distribution as a biasing distribution can result in very high-variance  estimates of the marginal and conditional likelihoods, which ultimately lead to large bias (and variance) when estimating the expected information gain. (We will demonstrate this issue in \cref{sec:numerical-results}.) To avoid this problem, a more sophisticated approach is needed to find biasing distributions that more closely resemble the joint and conditional posteriors.

\subsection{Sample sizes and scaling}
\label{ssec:scaling}
The nested estimator \eqref{eq:big-mc-estimator} is consistent (see \cite{rainforth2016nestedMC}) and in particular, its bias and variance converge to zero as $M_1, M_2,  N \to \infty$. Since all three sample sizes must increase, it is natural to ask whether there is an optimal allocation of inner ($M_1, M_2$) and outer ($N)$ samples for a given computational budget $W = N(M_1 + M_2)$.

For simplicity, let $M_1 = M_2 = M$ so that $W = 2MN$ and define $\widetilde{C} \coloneqq (C_1- C_2)^2>0$, where we drop dependence on $\vd$ to reduce clutter.  Let $\alpha^2=2M/N$ be the ratio between the number of inner and outer Monte Carlo samples. The value of $\alpha$ and the corresponding ratio $\alpha^2$ that minimize the leading-order MSE are
\begin{align}
    \alpha_\ast=2 \left ( \frac{\widetilde{C}}{D_3} \right )^{1/3} \frac{1}{W^{1/6}}
    \quad
    \Rightarrow\quad
    \alpha_\ast^2= 4 \left ( \frac{\widetilde{C}}{D_3} \right )^{2/3} \frac{1}{W^{1/3}}.
    \label{eq:work-alpha-opt}
\end{align}
This result suggests that the ratio between the number of inner and outer samples should decrease slowly as the computational budget increases, in particular scaling as $W^{-1/3}$. The ratio $\widetilde{C}/D_3$ reflects the relative magnitudes of leading-order terms in the bias and variance; if the bias dominates, one should choose a relatively larger number of inner loop samples. A detailed derivation of the optimal scaling and a numerical study of the asymptotic behavior are presented in \cref{appendix:scaling}.

It is important to emphasize, however, that all these results are asymptotic in $M, N \to \infty$. For finite $M$ and $N$, higher-order terms in the bias and variance may come into play. Also, the constants in \eqref{eq:work-alpha-opt} are in practice difficult to estimate, and an incorrect value---one that does not capture the relative magnitudes of the bias and the variance---may cause the $\alpha \sim W^{-1/6}$ scaling to perform worse than a fixed $M/N$ ratio for finite sample sizes. 

\subsection{Layered multiple importance sampling}
\label{ssec:lmis}

Now we introduce our new algorithm. 

\subsubsection{Overall algorithm}
To implement the focused EIG estimator \eqref{eq:big-mc-estimator}, one should identify good biasing distributions $q_\text{marg}^{(i)}$ and  $q_\text{cond}^{(i)}$ at each outer-loop iteration $i=1, \ldots, N$. An ideal approach might involve directly sampling from the posterior corresponding to each $\vy^{(i)}$, but doing so is computationally taxing---and, moreover, should only be done when evaluating the corresponding \emph{normalized} posterior density (to avoid instabilities of the harmonic mean estimator \cite{raftery1994}), which requires knowledge of the quantity we actually wish to estimate.
Instead, we will restrict our choice of biasing distributions to a parametric family with known density and use approximations of the joint and conditional posteriors, within this family, as biasing distributions. We describe our approach below using multivariate $t$-distributions; since these distributions are heavy-tailed, they are broadly useful in importance sampling \cite{mcbook}. As we shall emphasize later, however, our approach can easily be used with other parametric families.

To identify a multivariate $t$-distribution that is good for biasing relative to a given posterior, it is useful to estimate the posterior mean and covariance. Ideally, we would like to estimate these moments without drawing additional samples that require costly forward model evaluations. Our key idea is as follows: if we proceed sequentially  through the outer loop of \eqref{eq:big-mc-estimator}, indexed by $i$, then we can use the samples drawn from each inner-loop biasing distribution in the previous steps $\{1, \ldots, i-1\}$ to help characterize the posterior corresponding to the current $\vy^{(i)}$.
To reuse these existing samples, we consider a multiple importance sampling approach \cite{veach1995}, which combines the previously-drawn samples into a single mixture distribution 
that is used to estimate the mean $\vmu_\text{post}^{(i)}$ and covariance $\vSigma_\text{post}^{(i)}$ of the posterior distribution $p(\vtheta,\veta|\vy^{(i)},\vd)$. To this end, we first introduce self-normalized importance sampling estimators of the posterior mean and covariance:
\begin{align*}
    \widehat{\vmu}_{\text{post}}^{(i)}    & =\sum_{\ell=1}^{L} w_\text{post}^{(i,\ell)} \begin{bmatrix}\vtheta^{(i,\ell)}\\\veta^{(i,\ell)}\end{bmatrix}\,                                                                                                                                  \\[0.5em]
    \widehat{\vSigma}_{\text{post}}^{(i)} & =\sum_{\ell=1}^{L} w_\text{post}^{(i,\ell)} \biggl(\begin{bmatrix}\vtheta^{(i,\ell)}\\\veta^{(i,\ell)}\end{bmatrix}-\widehat{\vmu}_\text{post}^{(i)}\biggr)\biggl(\begin{bmatrix}\vtheta^{(i,\ell)}\\\veta^{(i,\ell)}\end{bmatrix}-\widehat{\vmu}_\text{post}^{(i)}\biggr)^\transpose
\end{align*}
where the samples ${\{(\vtheta^{(i,\ell)},\veta^{(i,\ell)})\}}_{\ell=1}^{L}$ are drawn from a biasing distribution $q_\text{post}^{(i)}(\vtheta,\veta)$ and the unnormalized importance weights are given by
\begin{align*}
    \widetilde{w}_\text{post}^{(i,\ell)} \propto \frac{\overbrace{p(\vtheta^{(i,\ell)},\veta^{(i,\ell)}|\vy^{(i)},\vd)}^{\text{joint posterior}}}{q_\text{post}^{(i)}(\vtheta^{(i,\ell)},\veta^{(i,\ell)})} \propto \frac{\overbrace{p(\vy^{(i)}|\vtheta^{(i,\ell)},\veta^{(i,\ell)},\vd)}^{\text{likelihood}}\,\overbrace{p(\vtheta^{(i,\ell)},\veta^{(i,\ell)})}^{\text{prior}}}{q_\text{post}^{(i)}(\vtheta^{(i,\ell)},\veta^{(i,\ell)})}, \label{eq:mis-weights-unnormalized}
\end{align*}
which can be rescaled to obtain the self-normalized importance weights
\begin{align}
    w_\text{post}^{(i,\ell)} = \frac{\widetilde{w}_\text{post}^{(i,\ell)}}{\sum_{\ell=1}^L\widetilde{w}_\text{post}^{(i,\ell)}}.
\end{align}

Now, let $q_\text{post}^{(i)}$ be a mixture distribution whose components are distributions from which we have already drawn samples---e.g., the prior distribution (since we can easily draw all $N$ outer loop samples in advance) and the inner-loop biasing distributions from earlier iterations:
\begin{align}
    q_\text{post}^{(i)}(\vtheta,\veta)=\frac{N}{L}p(\vtheta,\veta)+\frac{M_1}{L}q_\text{marg}^{(1)}(\vtheta,\veta) + \frac{M_1}{L}q_\text{marg}^{(2)}(\vtheta,\veta)+\cdots+\frac{M_1}{L}q_\text{marg}^{(i-1)}(\vtheta,\veta).
    \label{eq:mixturebias}
\end{align}
Instead of drawing new samples from this mixture distribution, we simply realize that the existing samples \emph{are} a draw from $q_\text{post}^{(i)}$:
\begin{align}
    \begin{split}
        \underbrace{{\{(\vtheta^{(i,\ell)},\veta^{(i,\ell)})\}}_{\ell=1}^{L}}_{\text{from $q_\text{post}^{(i)}$}} = \underbrace{{\{(\vtheta^{(i)},\veta^{(i)})\}}_{i=1}^N}_{\text{from prior}}
        &\sqcup
        \underbrace{{\{(\vtheta^{(1,j)},\veta^{(1,j)})\}}_{j=1}^{M_1}}_{\text{from $q_\text{marg}^{(1)}$}}
        \sqcup \underbrace{{\{(\vtheta^{(2,j)},\veta^{(2,j)})\}}_{j=1}^{M_1}}_{\text{from $q_\text{marg}^{(2)}$}} \\
        &\sqcup \dots \sqcup
        \underbrace{{\{(\vtheta^{(i-1,j)},\veta^{(i-1,j)})\}}_{j=1}^{M_1}}_{\text{from $q_\text{marg}^{(i-1)}$}}
    \end{split}
\end{align}
Moreover, because the forward model $\vG$ has already been evaluated each sample of $(\vtheta, \veta)$ in this empirical mixture, evaluating the likelihood at the current data point $\vy^{(i)}$ will be very cheap. This process results in the \emph{incremental enrichment} of the samples used to estimate the posterior mean $\vmu_\text{post}^{(i)}$ and covariance $\vSigma_\text{post}^{(i)}$. Pseudocode for this process is given in Algorithm~\ref{alg:mis}.

The inner estimators for the marginal and conditional likelihood still retain their original forms,~\eqref{eq:marginal-likelihood-estimator} and~\eqref{eq:conditional-likelihood-estimator}, but now the biasing distributions $q_\text{marg}^{(i)}$ and $q_\text{cond}^{(i)}$ are set according to the estimated posterior moments. Thus the biasing distribution for estimating the marginal likelihood is
\begin{align}
    q_\text{marg}^{(i)} (\vtheta,\veta) & =t_\nu(\widehat{\vmu}_\text{post}^{(i)}, \widehat{\vSigma}_\text{post}^{(i)}).
\end{align}
Since it is helpful for biasing distributions to have heavier tails, we typically choose small $\nu$, e.g., $\nu = 2.5$. One possible choice of $q_\text{cond}^{(i)}$ is another multivariate $t$-distribution derived from conditionals of the posterior Gaussian approximation:
\begin{align}
    q_\text{cond}^{(i)}(\veta) & =t_\nu\bigl(\vmu_{\veta}+\vSigma_{\vtheta\veta}^T\vSigma_{\vtheta}^{-1}(\vtheta^{(i)}-\vmu_{\vtheta}), \vSigma_{\veta}-\vSigma_{\vtheta\veta}^T\vSigma_{\vtheta}^{-1}\vSigma_{\vtheta\veta}\bigr).
    \label{eq:qcondsimple}
\end{align}
where we used the block-matrix notation
\begin{align*}
    \widehat{\vmu}_\text{post}^{(i)} & = \begin{bmatrix}\vmu_{\vtheta}\\\vmu_{\veta}\end{bmatrix},\quad
    \widehat{\vSigma}_\text{post}^{(i)} = \begin{bmatrix}\vSigma_{\vtheta}&\vSigma_{\vtheta\veta}\\\vSigma_{\vtheta\veta}^T&\vmu_{\veta}\end{bmatrix}.
\end{align*}
Alternatively, one could use the conditional of $q_\text{marg}^{(i)}$ directly; note that this is just another multivariate $t$ distribution, very similar to \eqref{eq:qcondsimple}, but with a scaling of the scale matrix $\vSigma_{\veta \vert \vtheta} \coloneqq \vSigma_{\veta}-\vSigma_{\vtheta\veta}^T\vSigma_{\vtheta}^{-1}\vSigma_{\vtheta\veta}$ and a modification of the degrees of freedom $\nu$ \cite{ding2016conditional}. Another possible choice is simply the $\veta$--marginal of  $q_\text{marg}^{(i)}$. This last choice is more diffuse, since it does not depend on $\vtheta^{(i)}$, but may be useful in situations where the preceding approximations to the true posterior conditional are poor.

Note, finally, that the importance sampling estimates for the marginal and conditional likelihood involve drawing \emph{new} samples $\{(\vtheta^{(i,j)},\veta^{(i,j)})\}_{j=1}^{M_1}$ and $\{(\veta^{(i,k)})\}_{k=1}^{M_2}$ from the biasing distributions just constructed. Since we know the normalized densities of these biasing distributions, the resulting importance sampling estimates are unbiased at finite $M_1, M_2$. 

We call this complete strategy
\emph{layered multiple importance sampling}, with two layers:
\begin{enumerate}
    \item A biased layer (with bias due to self-normalization) to estimate posterior moments, via multiple importance sampling with sample reuse;
    \item An unbiased layer to estimate the marginal and conditional likelihoods, using biasing distributions selected from a parametric family according to the estimated posterior moments.
\end{enumerate}

\subsubsection{Pruning the mixture biasing distribution}
Although layered multiple importance sampling does not require additional model evaluations beyond a standard nested scheme, if we include all the biasing distributions from earlier iterations in \eqref{eq:mixturebias}, the number of likelihood evaluations---using cached model evaluations---grows as $O(N^2M_1^2)$, which can result in a nontrivial computational overhead. %
Instead, we can include only the most ``useful'' samples in the empirical mixture distribution. Since we do not know the posterior beforehand, we can use the $(\vtheta^{(i)},\veta^{(i)})$ which gave rise to the data $\vy^{(i)}$ as our best guess for where the posterior $p(\vtheta,\veta|\vy^{(i)},\vd)$ is most concentrated. This selection criterion allows us to define an index set
\begin{align}
    \mathcal{J}^{(i)}=\{m\,:\, q_\text{marg}^{(m)}(\vtheta^{(i)},\veta^{(i)}) > p(\vtheta^{(i)},\veta^{(i)}),\, m<i\}.
    \label{eq:pruning-criterion}
\end{align}
where we only include distributions that have a higher probability density on the sample $(\vtheta^{(i)},\veta^{(i)})$ than the prior. Then, the mixture density \eqref{eq:mixturebias} can be modified as
\begin{align}
    q_\text{post}^{(i)}(\vtheta,\veta) = \frac{N}{L} p(\vtheta,\veta) + \frac{M_1}{L}  \sum_{m\in\mathcal{J}^{(i)}} q_\text{marg}^{(m)}(\vtheta,\veta),
\end{align}
and the samples drawn from this mixture distribution %
$q_\text{post}^{(i)}$ can be written as a disjoint union of samples drawn previously from the prior and biasing distributions:
\begin{align}
    \underbrace{\{(\vtheta^{(i,\ell)},\veta^{(i,\ell)})\}_{\ell=1}^{L}}_{\text{from $q_\text{post}^{(i)}$}} = \underbrace{\{(\vtheta^{(i)},\veta^{(i)})\}_{i=1}^N}_{\text{from prior}} \sqcup \bigsqcup_{m\in\mathcal{J}^{(i)}} \underbrace{\{(\vtheta^{(i,m,j)},\veta^{(i,m,j)})\}_{j=1}^{M_1}}_{\text{from $q_\text{marg}^{(m)}$}},
\end{align}
where the total number of samples is
\begin{align}
    L=N+M_1|\mathcal{J}^{(i)}|.
\end{align}
The specification of an index set $\mathcal{J}^{(i)}$ allows some generality: instead of the strategy in \eqref{eq:pruning-criterion}, one could include all the mixture components by setting $\mathcal{J}^{(i)}=\{1,2,\dots,i-1\}$, or one could avoid incremental enrichment by setting $\mathcal{J}^{(i)}=\{ \emptyset \}$, so that $q_\text{post}^{(i)}$ becomes the prior distribution $p(\vtheta,\veta)$ and we reuse only the $N$ samples drawn from the prior.

\begin{figure}[htbp]
    \centering
    \includegraphics[width=5in]{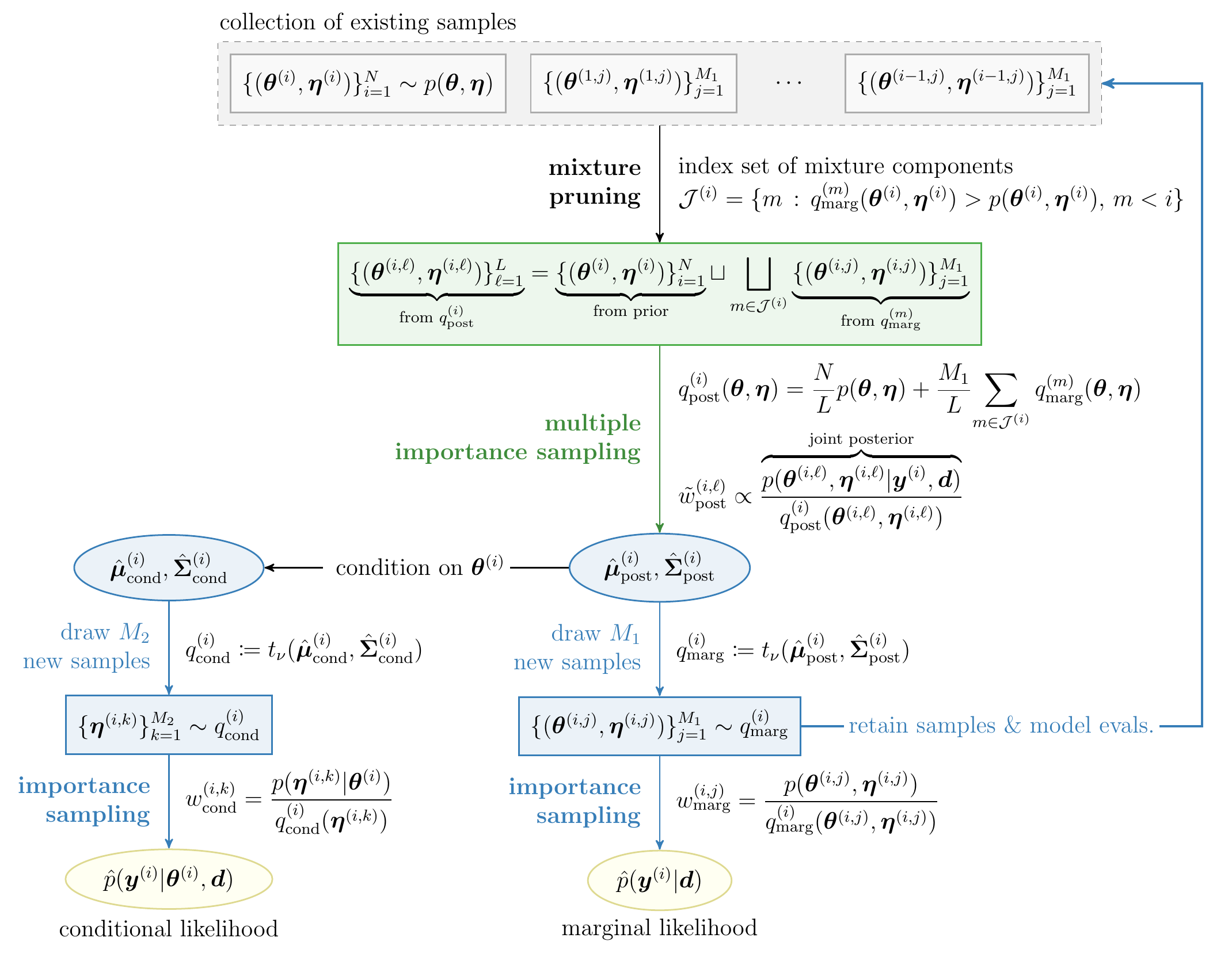}
    \caption{Schematic illustrating the layered multiple importance sampling scheme.}
    \label{fig:flowchart}
\end{figure}

Our overall strategy---layered multiple importance sampling with mixture pruning---is illustrated via the flowchart in \cref{fig:flowchart}.

\subsubsection{Ordering samples}
Since the layered multiple importance sampling approach is sequential by construction, a natural question is whether the ordering of the prior samples $\{(\vtheta^{(i)},\veta^{(i)},\vy^{(i)})\}_{i=1}^{N}$ will affect the behavior of the overall estimator. Since we draw these samples before evaluating the inner estimators, we have the freedom to specify a particular ordering. Intuitively, a good ordering is one that maximizes incremental enrichment---such that samples drawn from earlier biasing distributions will be reused in many subsequent mixture distributions. This goal suggests an ordering that prioritizes the posterior distributions $p(\vtheta,\veta|\vy^{(i)},\vd)$ that overlap the most with all other posterior distributions. However, since the $N$ different posterior distributions are not characterized before the ordering is chosen, we suggest using the prior density of $p(\vtheta^{(i)},\veta^{(i)})$ as a surrogate. With sufficient data and under appropriate consistency conditions, the posterior distributions $p(\vtheta,\veta|\vy^{(i)},\vd)$ will be concentrated near the $\vtheta^{(i)},\veta^{(i)}$ that gave rise to the $\vy^{(i)}$; more generally, the centers of the posteriors will (collectively) be distributed similarly to the prior. Thus, ordering the samples by decreasing prior density should achieve a similar effect as ordering the posteriors by decreasing overlap.

Our complete algorithm, LMIS with mixture pruning and an initial sorting of samples, is described in Algorithm~\ref{alg:adaptalg}.

\begin{figure}[htbp]
    \centering
    \includegraphics[width=2.7in,trim={0 0 0 0},clip]{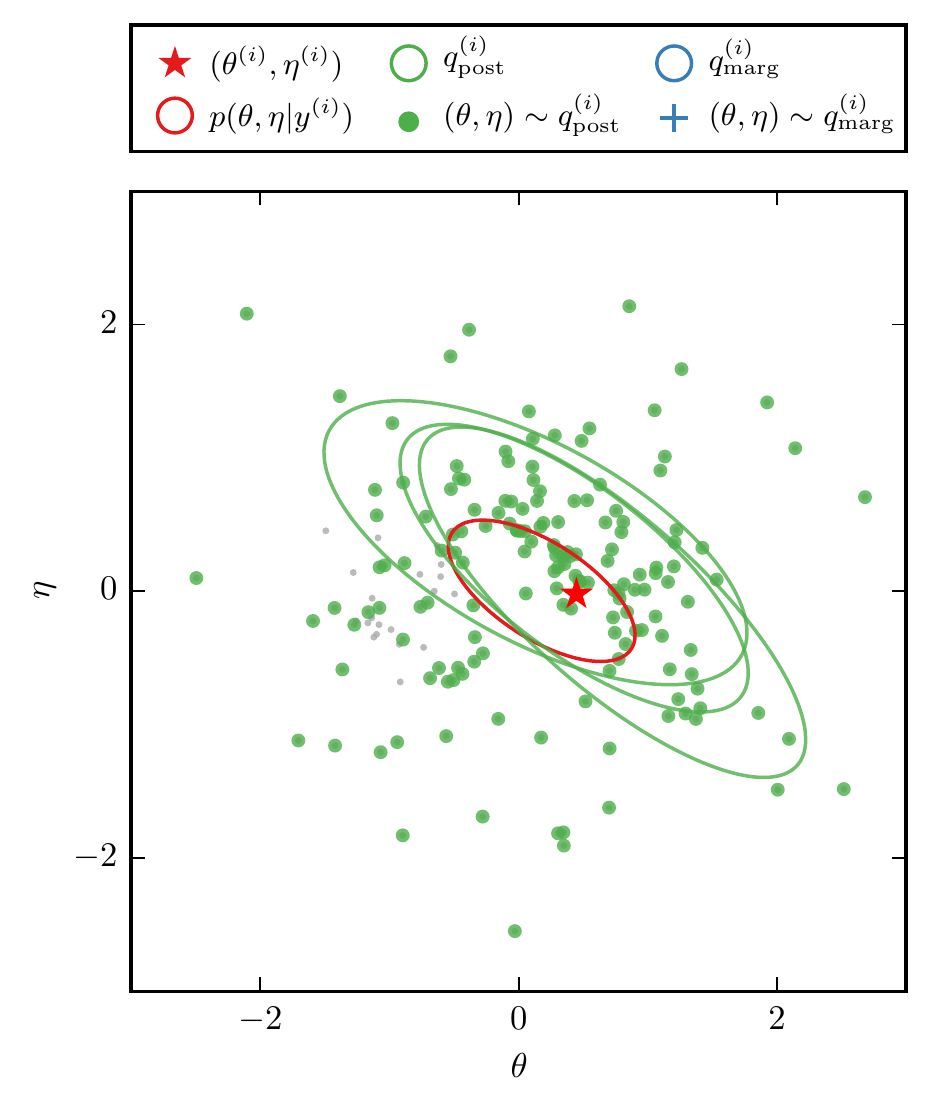}
    \includegraphics[width=2.5in,trim={0.7cm 0 0 1.6cm},clip]{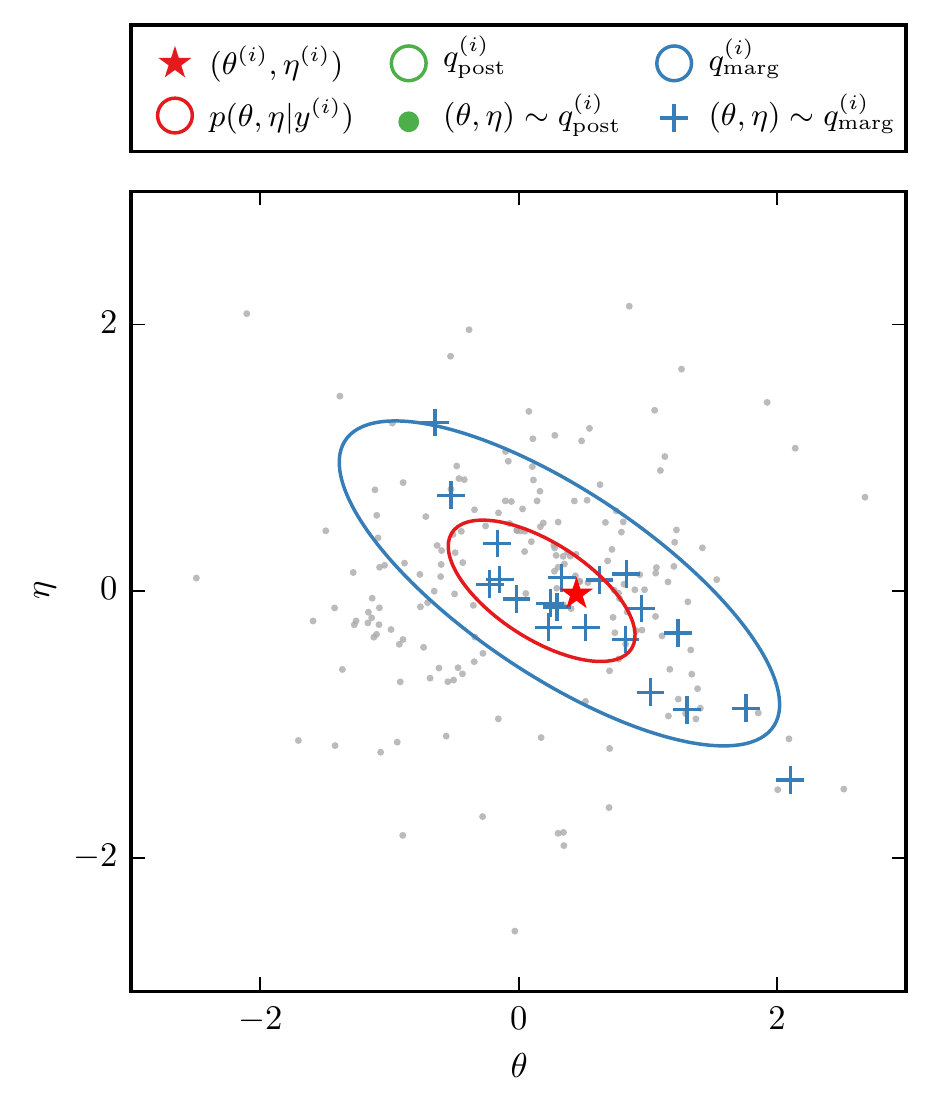}
    \caption{Illustration of layered multiple importance sampling on a two-dimensional linear Gaussian example $(k=1, d=0.5, \sigma_\epsilon=0.4)$ with $N=100$, $M_1=M_2=20$ samples. The posterior is drawn in \textcolor{red}{red}, components of the mixture biasing distribution in \textcolor{green!50!black}{green}, and the biasing distribution used to estimate the marginal likelihood in \textcolor{blue!50!gray}{blue}. Existing samples not involved in estimating the marginal likelihood are drawn in \textcolor{gray!50!black}{gray}.}
    \vspace{-1em}
    \label{fig:lmis-schematic}
\end{figure}

\begin{algorithm2e}[htbp]
    \DontPrintSemicolon

    \SetKwProg{func}{Function}{}{}
    \SetKwFunction{EIG}{ExpectedInformationGain}
    \SetKwFunction{MISMoments}{EstimatePosteriorMoments}
    \SetKwProg{func}{Function}{}{}
    \SetFuncSty{textsl}

    \func{\EIG{$\vd$}}{
    \For{$i \gets 1 \ldots N$}{
        Sample $(\vtheta^{(i)},\veta^{(i)})$ from  the prior $p(\vtheta,\veta)$\;
        Sample $\vy^{(i)}$ from the likelihood $p(\vy|\vtheta^{(i)},\veta^{(i)},\vd)$\;
    }
    Rearrange $\{(\vtheta^{(i)},\veta^{(i)})\bigr\}_{i=1}^N$ and $\vy^{(i)}$ in decreasing order of $p(\vtheta^{(i)},\veta^{(i)})$ \;
    \For{$i \gets 1 \ldots N$}{
    $\mathcal{J}^{(i)}\gets \{m:q_\text{marg}^{(m)}(\vtheta^{(i)},\veta^{(i)}) > p(\vtheta^{(i)},\veta^{(i)}),\,m<i\}$\;
    $\widehat{\vmu}_\text{post}^{(i)},\,\widehat{\vSigma}_\text{post}^{(i)} \gets \,$\MISMoments{$\vy^{(i)}$, $\mathcal{J}^{(i)}$}\;
    \BlankLine
    \tcp{Estimate the marginal likelihood $p\lparen\vy^{(i)}|\vd\rparen$}
    $q_\text{marg}^{(i)}\gets t_\nu(\widehat{\vmu}_\text{post}^{(i)},\widehat{\vSigma}_\text{post}^{(i)})$\;
    \BlankLine
    \For{$j\gets 1\ldots M_1$}{
    Sample $(\vtheta^{(i,j)}, \veta^{(i,j)})$ from $q_\text{marg}^{(i)}(\vtheta^{(i,j)}, \veta^{(i,j)})$\;
    }
    $\displaystyle \widehat{p}(\vy^{(i)}|\vd)\gets \frac{1}{M_1}\sum_{j=1}^{M_1}p(\vy^{(i)}|\vtheta^{(i,j)},\veta^{(i,j)},\vd)\,{p(\vtheta^{(i,j)},\veta^{(i,j)})}/{q_\text{marg}^{(i)}(\vtheta^{(i,j)},\veta^{(i,j)})}$\;
    \BlankLine
    \tcp{Estimate the conditional likelihood $p\lparen\vy^{(i)}|\vtheta^{(i)}, \vd\rparen$}
    Condition $\widehat{\vmu}_\text{post}^{(i)}$ and $\widehat{\vSigma}_\text{post}^{(i)}$ on $\vtheta^{(i)}$ to obtain $\widehat{\vmu}_\text{cond}^{(i)}$ and $\widehat{\vSigma}_\text{cond}^{(i)}$\;%
    $q_\text{cond}^{(i)}\gets t_\nu(\widehat{\vmu}_\text{cond}^{(i)},\widehat{\vSigma}_\text{cond}^{(i)})$\;
    \BlankLine
    \For{$k \gets 1 \ldots M_2$}{ 
        Sample $\veta^{(i,k)}$ from $q_\text{cond}^{(i)}(\veta)$\;
    }
    $\displaystyle \widehat{p}(\vy^{(i)}|\vtheta^{(i)},\vd)\gets\frac{1}{M_2}\sum_{k=1}^{M_2}p(\vy^{(i)}|\vtheta^{(i)},\veta^{(i,k)},\vd)\,{p(\veta^{(i,k)})}/{q_\text{cond}^{(i)}(\veta^{(i,k)})}$\;
    }
    $\displaystyle \widehat{U}(\vd)\gets \frac{1}{N}\sum_{i=1}^N\log\bigl[\widehat{p}(\vy^{(i)}|\vtheta^{(i)},\vd)\bigr]-\log\bigl[\widehat{p}(\vy^{(i)}|\vd)\bigr]$
    }
    \caption{Layered multiple importance sampling estimate of the expected information gain}
    \label{alg:adaptalg}
\end{algorithm2e}

\begin{algorithm2e}[htbp]
    \setstretch{1.25}
    \DontPrintSemicolon
    \SetFuncSty{textsl}
    \SetKwProg{func}{Function}{}{}
    \SetKwFunction{EIG}{ExpectedInformationGain}
    \SetKwFunction{MISMoments}{EstimatePosteriorMoments}
    \SetKwProg{func}{Function}{}{}
    \SetFuncSty{textsl}

    \func{\MISMoments{$\vy^{(i)}$, $\mathcal{J}^{(i)}$}}{

    \tcp{Create mixture biasing distribution $q_\text{post}^{(i)}$ according to the index set $J^{(i)}$}
    Initialize $\bigl\{(\vtheta^{(i,\ell)},\veta^{(i,\ell)})\bigr\}_{\ell=1}^{N}\gets \{(\vtheta^{(i)},\veta^{(i)})\bigr\}_{i=1}^N$ drawn earlier from the prior \;
    $L\gets N$\;
    \For{$m\in\mathcal{J}^{(i)}$}{
    $\bigl\{(\vtheta^{(i,\ell)},\veta^{(i,\ell)})\bigr\}_{\ell=1}^{L+M_1} \gets \bigl\{(\vtheta^{(i,\ell)},\veta^{(i,\ell)})\bigr\}_{\ell=1}^{L}\sqcup \bigl\{(\vtheta^{(m,j)}, \veta^{(m,j)})\bigr\}_{j=1}^{M_1}$ %
    \;
    $L\gets L+M_1$
    }

    $\displaystyle q_\text{post}^{(i)}(\vtheta,\veta)\gets \frac{N}{L} p(\vtheta,\veta) + \frac{M_1}{L}\sum_{m\in\mathcal{J}^{(i)}}q_\text{marg}^{(m)}(\vtheta,\veta)$\;
    \tcp{Compute importance weights and normalize them}
    \For{$(\vtheta^{(i,\ell)},\veta^{(i,\ell)})\in\bigl\{(\vtheta^{(i,\ell)},\veta^{(i,\ell)})\bigr\}_{\ell=1}^{L}$}{
    \BlankLine
    $\widetilde{w}_\text{post}^{(i,\ell)}\gets \dfrac{p(\vy^{(i)}|\vtheta^{(i,\ell)},\veta^{(i,\ell)},\vd)\,p(\vtheta^{(i,\ell)},\veta^{(i,\ell)})}{q_\text{post}^{(i)}(\vtheta^{(i,\ell)},\veta^{(i,\ell)})}$
    }
    $w_\text{post}^{(i,\ell)} \gets \widetilde{w}_\text{post}^{(i,\ell)} / \sum_{\ell=1}^L \widetilde{w}_\text{post}^{(i,\ell)}$\;
    \tcp{Estimate posterior mean and covariance}
    $\widehat{\vmu}_{\text{post}}^{(i)} \gets \sum_{\ell=1}^{L}w_\text{post}^{(i,\ell)}\Bigl[\begin{smallmatrix}\vtheta^{(i,\ell)}\\\veta^{(i,\ell)}\end{smallmatrix}\Bigr]\, $\;
    $\widehat{\vSigma}_{\text{post}}^{(i)} \gets \sum_{\ell=1}^{L}w_\text{post}^{(i,\ell)}\Bigl(\Bigl[\begin{smallmatrix}\vtheta^{(i,\ell)}\\\veta^{(i,\ell)}\end{smallmatrix}\Bigr]-\widehat{\vmu}_\text{post}^{(i)}\Bigr)\Bigl(\Bigl[\begin{smallmatrix}\vtheta^{(i,\ell)}\\\veta^{(i,\ell)}\end{smallmatrix}\Bigr]-\widehat{\vmu}_\text{post}^{(i)}\Bigr)^\transpose $\;
    }
    \caption{Multiple importance sampling estimate of posterior moments}
    \label{alg:mis}
\end{algorithm2e}

\section{Numerical results}%
\label{sec:numerical-results}
In this section, we examine two experimental design problems. The first is a linear Gaussian problem with a known, exact solution; this example is used to assess the performance and convergence properties of the layered multiple importance sampling algorithm, for varying dimensions. The second example is a more challenging nonlinear experimental design problem that occurs in M\"ossbauer spectroscopy, where we again illustrate the computational efficieny of the LMIS approach as well as the impact of the focused design objective.

\subsection{Higher-dimensional linear Gaussian example}%
\label{ssec:hd-linear-gaussian}
We revisit the linear Gaussian problem that was first introduced in \cref{ssec:linear-gaussian}, but increase the dimension of the inference problem to $n=4$ and $n=8$ dimensions. The dimension of the parameter of interest remains the same, i.e., $n_\theta=1$, so the dimension of the nuisance parameters is $n_\eta = n-n_\theta$. 
In general, the observation model for our $n$-dimensional inference problem is:
\begin{align}
    \begin{bmatrix}
        y_1    \\
        y_2    \\
        \vdots \\
        y_{n}
    \end{bmatrix}=
    \underbrace{\begin{bmatrix}
            kd &        &        & 1      \\
               & k(1-d) &        &        \\
               &        & \ddots &        \\
            1  &        &        & k(1-d)
        \end{bmatrix}}_{\vG(d,k)}
    \begin{bmatrix}\theta\\\eta_1\\\vdots\\\eta_{n-1}\end{bmatrix} +
    \begin{bmatrix}\epsilon_1\\\epsilon_2\\\vdots\\\epsilon_{n}\end{bmatrix}, \quad k>0.
    \label{eq:linnd}
\end{align}
The design matrix $\vG$ is parameterized by the parameter $d\in[0,1]$ and a gain parameter $k>0$.
The diagonal terms describe a trade-off between learning about the parameter of interest and learning about the nuisance parameters where $d=1$ maximizes the information gain in $\theta$ at the expense of information gain in the nuisance parameters $(\eta_1,\eta_2,\dots,\eta_{n-1})$. Furthermore, the nonzero off-diagonal terms in the design matrix introduce coupling between the parameter of interest and the nuisance parameters so that the resulting optimal design is not trivially $d=1$.

Each of the input parameters have standard normal priors, i.e., $(\theta,\eta_1,\dots,\eta_{n-1})^T\sim\mathcal{N}(\boldsymbol{0}, \vGamma^\text{pr})$, with $\vGamma^\text{pr}=\mathbf{I}_{n \times n}$.
The observation noise is additive Gaussian with zero mean and variance $\sigma_\epsilon^2$. i.e., $\vepsilon\sim\mathcal{N}(\boldsymbol{0},\vGamma^\text{obs})$ where $\vGamma^\text{obs}=\sigma_\epsilon^2\mathbf{I}_{n \times n}$.

The EIG for all of the parameters is
\begin{align*}
	U_\text{joint}(d)=-\frac{1}{2} \log \bigl[\det(\vGamma^\text{post})\bigr] 
\end{align*}
where $\vGamma^\text{post}$ is the posterior covariance matrix, given by
\begin{align*}
	\vGamma^\text{post}=\vGamma^\text{pr}-\vGamma^\text{pr}\vG^T(\vG\vGamma^\text{pr}\vG^T+\vGamma^\text{obs})^{-1}\vG\vGamma^\text{pr},
\end{align*}
and the EIG for the marginal in the parameter of interest is
\begin{align*}
	U_\text{marg}(d)=-\frac{1}{2} \log \bigl[\gamma^{\text{post}}_{1,1}\bigr]
\end{align*}
where $\gamma^{\text{post}}_{1,1}$ is the  top left entry of the posterior covariance matrix $\vGamma^\text{post}$.

\subsubsection{Expected information gain profiles}
\begin{figure}[htbp]
\centering
	\includegraphics[width=0.9\textwidth]{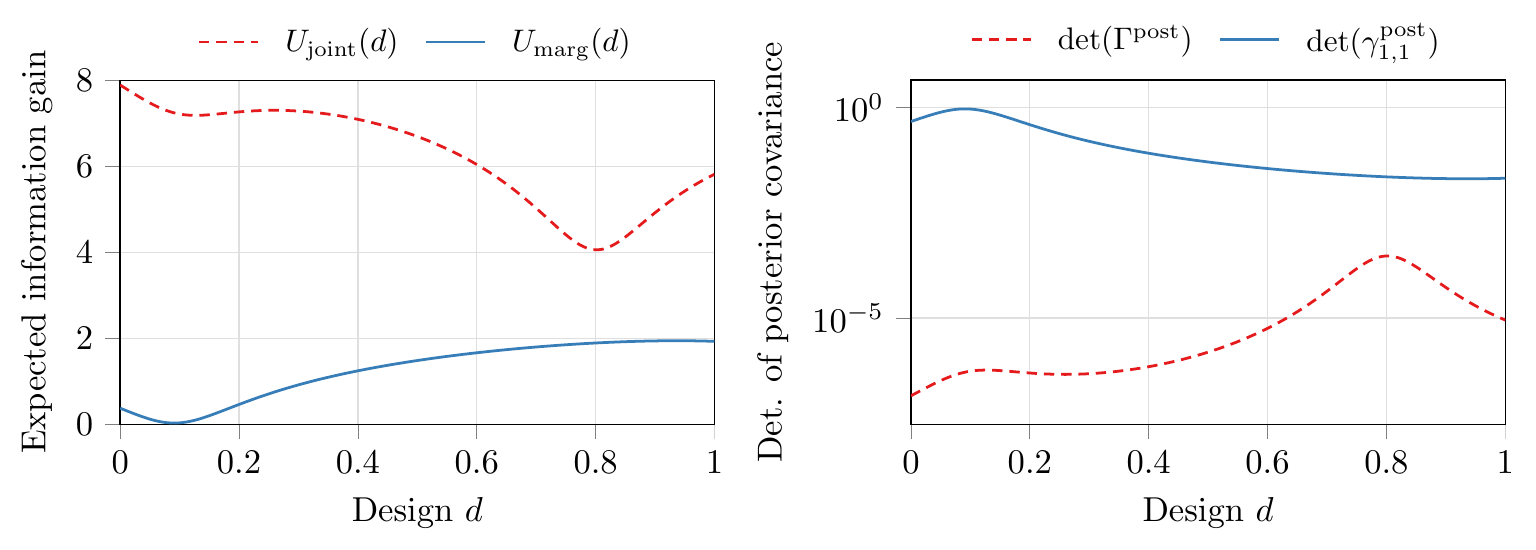}%
	\caption{(Left) EIG for focused (marginal) and unfocused (joint) objectives. (Right) Determinant of the posterior covariance matrix as a function of the design $d$.}
	\label{fig:eig-exact}
\end{figure}

For the following numerical examples in $n=4$ and $n=8$ dimensions, we set $k=5$ and $\sigma_\epsilon=0.4$.
The EIG for the focused and unfocused settings are plotted in \cref{fig:eig-exact}.
We see that the optimal designs for the two objectives are very different: the optimal design for learning about all the parameters is $d_\text{joint}^\ast=0$ and the optimal design for learning only about the parameter of interest is $d_\text{marg}^\ast\approx 0.93$.
The determinant of the posterior covariance matrix is plotted in \cref{fig:eig-exact} as a function of the design parameter $d$, which illustrates the relationship between the posterior covariance and the EIG.

\begin{figure}[htbp]
	\centering
	\subfloat[Expected value of $\widehat{U}_\text{joint}(d)$]{\label{fig:eig-joint}%
		\includegraphics[width=0.45\textwidth]{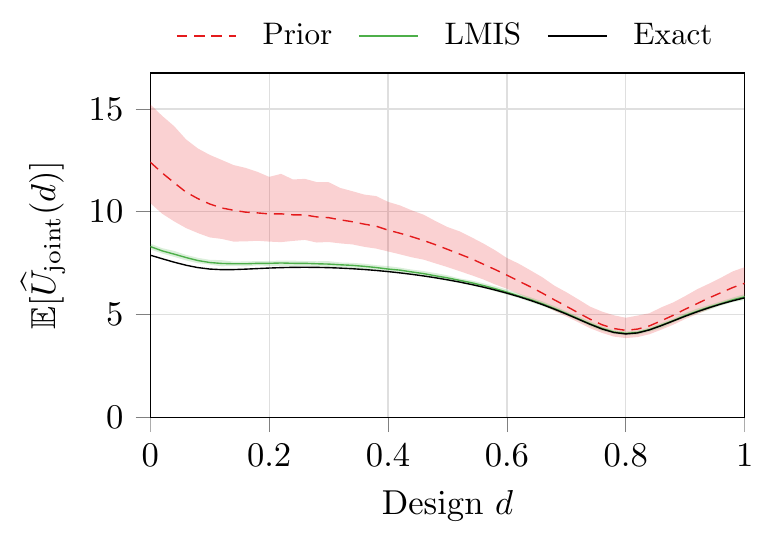}%
	}%
	\subfloat[Expected value of $\widehat{U}_\text{marg}(d)$]{\label{fig:eig-marginal}%
		\includegraphics[width=0.45\textwidth]{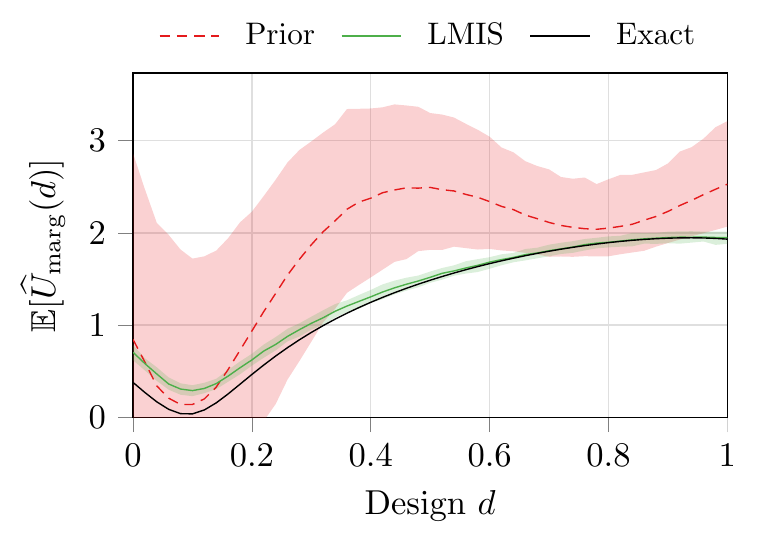}%
	}
	\caption{Estimated EIG profile over the design space for the 4D linear Gaussian example.
		Shaded areas represent the interval containing 95\% of $2000$ independent estimates of EIG at each $d$, for different sampling schemes. Red dashed and solid green lines are the means of these estimates.
		The ratios of the number of outer and inner Monte Carlo samples were selected to be favorable for each scheme according to \cref{fig:mse-ratio}, i.e., $N=500$, $M_1=M_2=50$ (LMIS) and $N=50$, $M_1=M_2=500$ (prior)}%
\end{figure}
From the analysis of the EIG estimator in \cref{eq:delta-bias} and \cref{eq:delta-variance}, we expect that the bias and variance will be largest when the biasing distributions and the posterior distributions have the smallest overlap.
Thus, if we use the the prior distribution as the biasing distribution, the bias and variance of the EIG estimator should be large when there is significant posterior concentration.
Accordingly, in \cref{fig:eig-exact} we observe that the volume (computed as the determinant of the posterior covariance matrix) of the posterior distribution varies over several orders of magnitude across the design space, and in \cref{fig:eig-joint} and \cref{fig:eig-marginal} we observe large positive bias for designs that correspond to significant posterior concentration.

The correlation of estimator bias with posterior concentration is particlularly undesirable behavior---particularly in the context of focused experimental design, since bias that is correlated with the determinant of the joint posterior can mask the true shape of the objective function (which depends only on the marginal posterior).
We observe this behavior in \cref{fig:eig-marginal} where the location of local and global maxima correspond to sub-optimal designs that maximize posterior concentration instead of those that maximize information gain in the parameter of interest.
Furthermore, when using the prior as a biasing distribution, the large estimator variance may pose a challenge for stochastic optimization algorithms. On the other hand, LMIS yields estimates of the EIG with much smaller bias and variance while using the same number of model evaluations.
More importantly, the LMIS estimates of the EIG in both the focused and unfocused cases capture the correct locations of the global maxima.

\begin{figure}[htbp]
	\centering
	\includegraphics[width=0.9\textwidth]{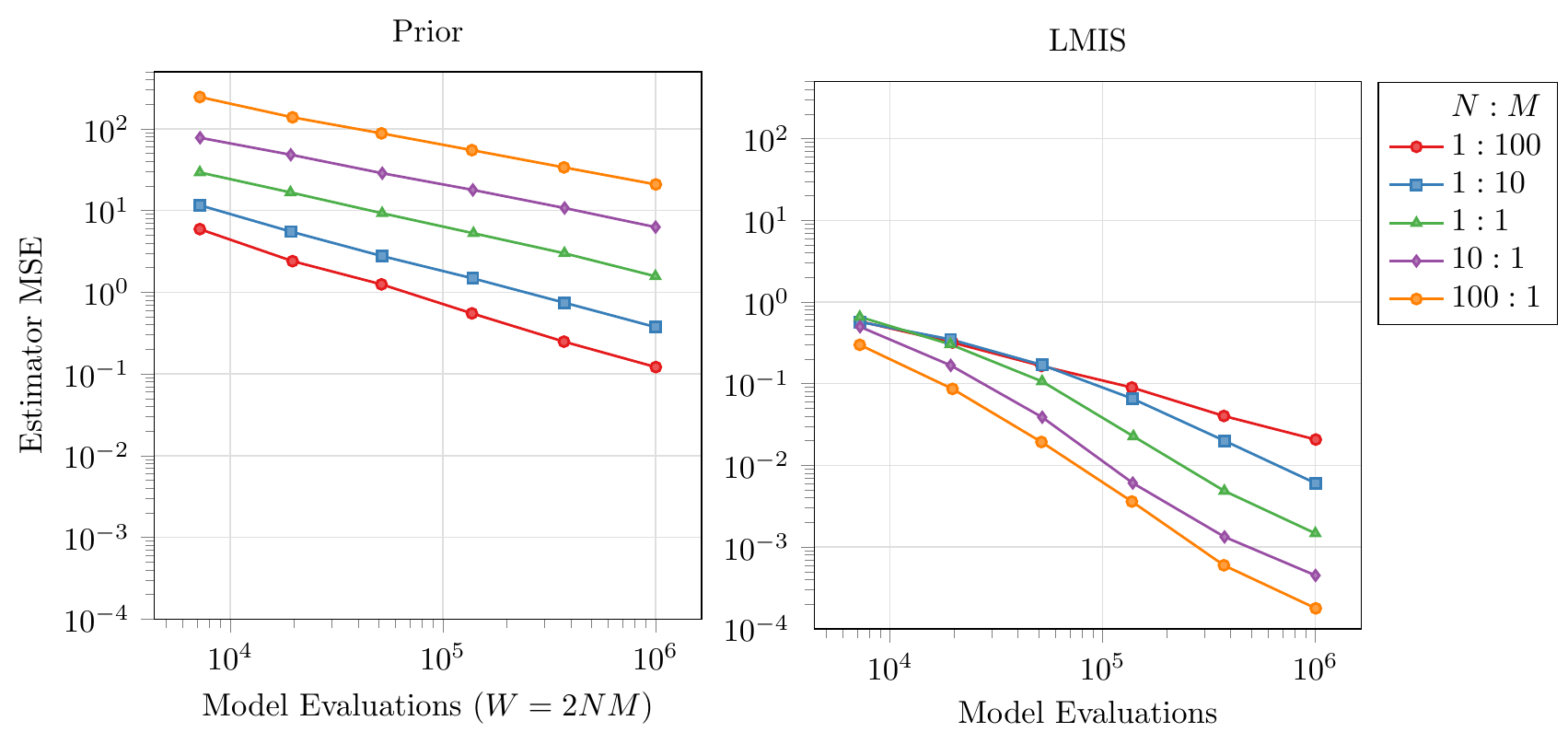}
	\caption{Mean squared error of the EIG estimator versus the number of model evaluations, for different ratios between the number of outer and inner samples.}
	\label{fig:mse-ratio}
\end{figure}

\subsubsection{Sample allocations, bias, and variance}
The allocation of samples between the outer and inner estimators can play an important role in the bias and variance of the overall estimator.
In \cref{fig:mse-ratio}, we investigate the effect of the ratio between the number of outer and inner samples on the mean squared error of the EIG estimator, both using the prior as a biasing distribution and using layered multiple importance sampling.
When using the prior as a biasing distribution, the bias dominates the mean squared error, so allocating more inner samples---at the expense of affording fewer outer samples---will reduce the overall mean squared error.
On the other hand, with LMIS, the biasing distributions closely approximate the posterior distributions, so fewer inner samples are required, and more samples can be allocated for the outer estimator, which not only decreases the overall variance, but also provides additional opportunities for incremental enrichment, further improving the approximation accuracy of the inner estimators.

\begin{figure}[htbp]
	\centering
	\includegraphics[width=5.45in]{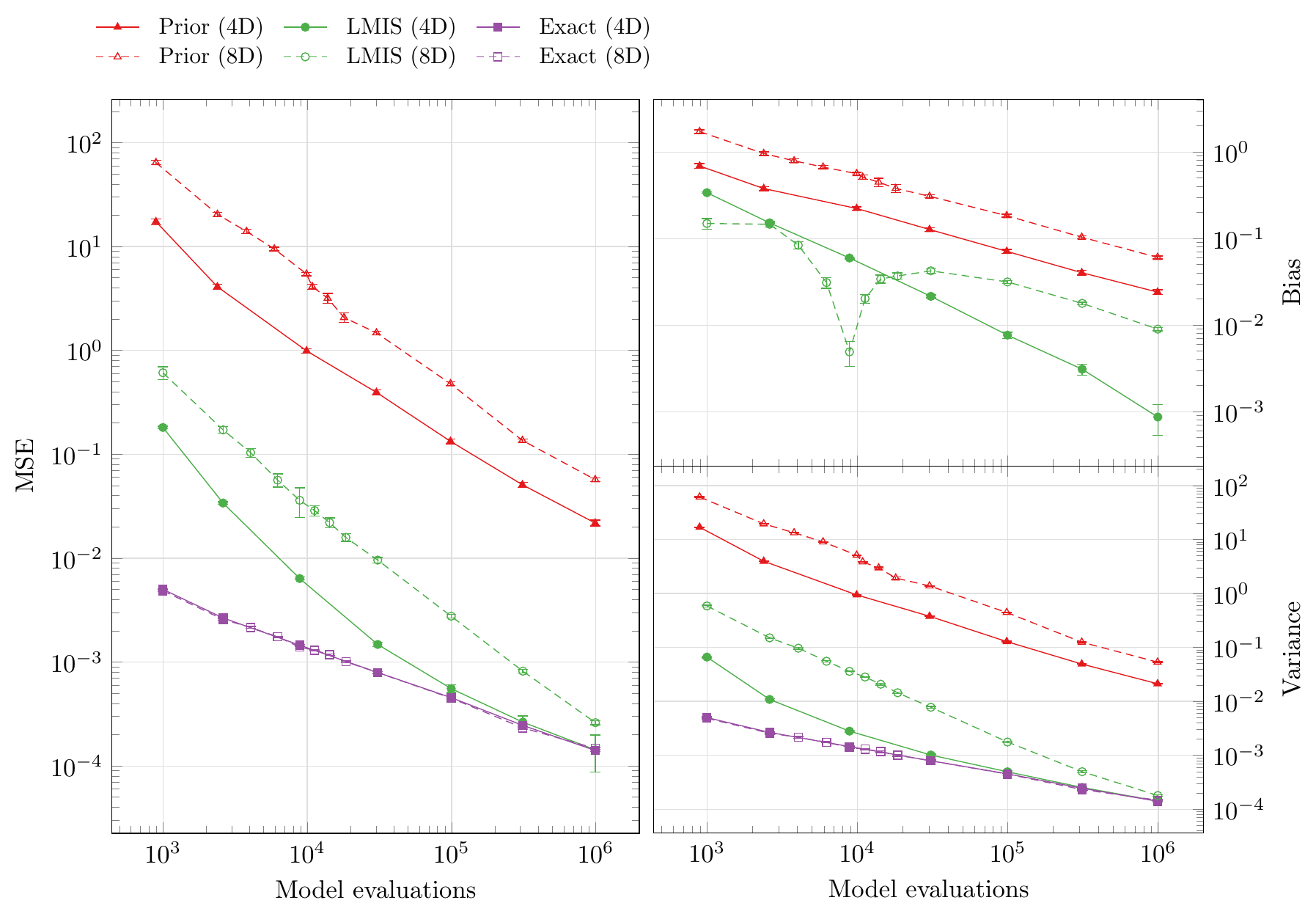}
	\caption{(Left) Mean squared error of the EIG estimator. `Exact' refers to using the analytical posterior distribution as the biasing distribution. (Right) Mean squared error decomposed into the variance and the absolute value of the bias. Error bars denote the Monte Carlo standard errors of these bias and variance estimates, obtained by averaging over replicates. The dip in estimator bias when using LMIS in the 8-dimensional case is due to taking the absolute value of the bias when the bias changes sign.}
	\label{fig:mse-convergence}
\end{figure}

Using favorable but fixed ratios between the number of outer and inner samples,\footnote{When using prior-based sampling, we set $M=100N$ and when using LMIS, we set $N=100M$.} we can examine the mean squared error as a function of number of model evaluations.
In \cref{fig:mse-convergence}, we observe a dramatic reduction in the mean squared error when using LMIS. 
With $10^6$ model evaluations, the resulting LMIS estimate has mean squared error comparable to using the exact posteriors as biasing distributions.\footnote{When using the exact joint and conditional posteriors as biasing distributions, the importance estimators of the marginal and conditional likelihoods are zero-variance estimators. In theory, a single inner sample is then sufficient to obtain an exact value of the marginal and conditional likelihood. However, for a fair comparison, we use the same number of inner and outer samples as LMIS, even if it is not the most efficient allocation when using the exact posterior.}
\begin{figure}[htbp]\centering
	\includegraphics[width=4.7in]{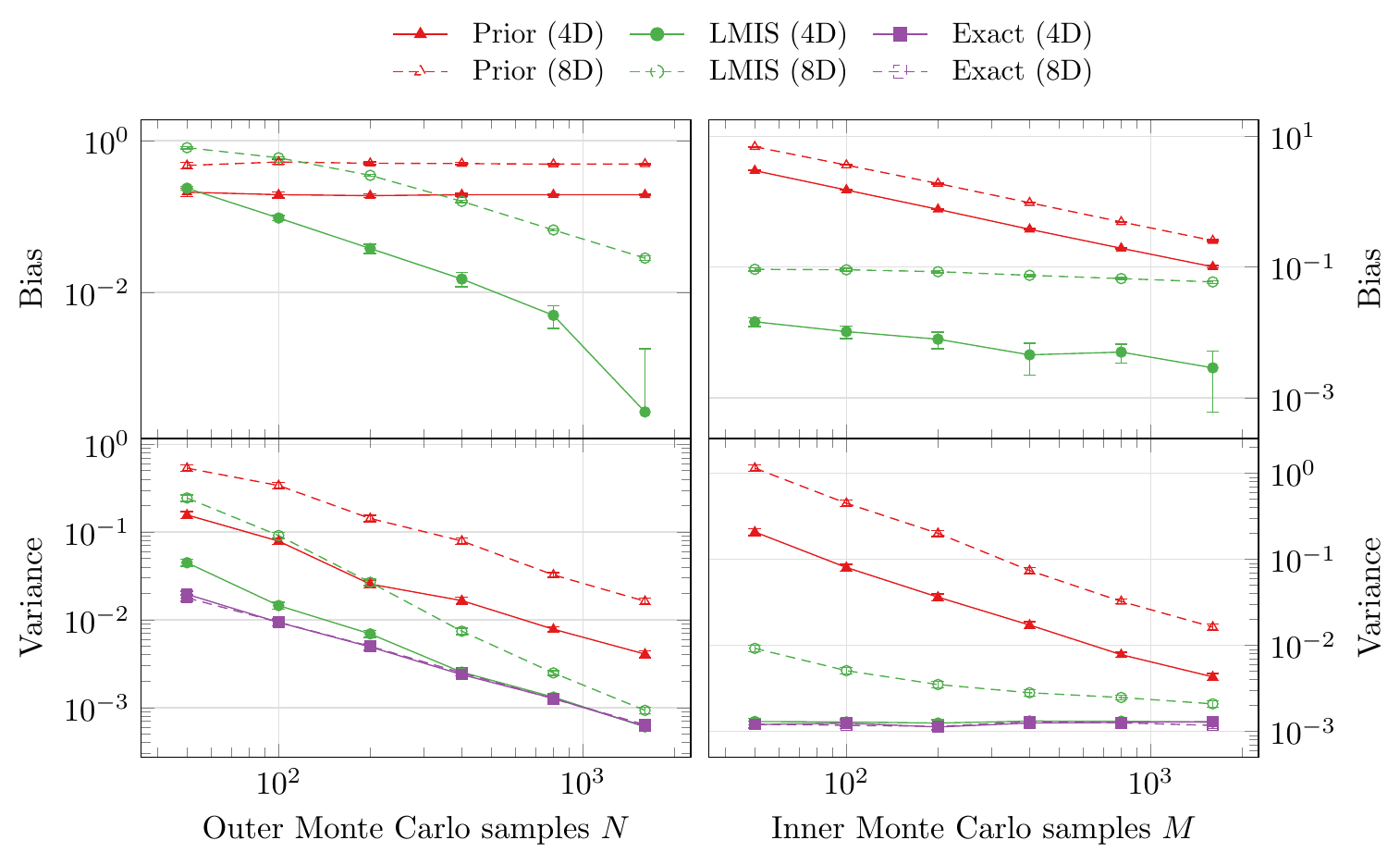}
	\caption{Bias and variance of the EIG estimators with fixed numbers of inner and outer samples. On the left,  the outer loop sample size $N$ is varied while the inner loop sample size $M=800$ is held fixed. On the right $M$ is varied while $N=800$ is held fixed.}
	\label{fig:convergence_sweep}
\end{figure}
To better understand the observed convergence rates in \cref{fig:mse-convergence}, we plot the mean squared error versus the number of inner and outer samples in \cref{fig:convergence_sweep}.
When using the prior as the biasing distribution, we observe the convergence rates predicted by \cref{eq:delta-bias} and \cref{eq:delta-variance}. The estimator bias decreases as $1/M$ and is unaffected by the number of outer samples, $N$. The variance also decreases with increasing number of inner and outer samples, which suggests that the $1/NM$ terms in \cref{eq:delta-variance} are not negligible.

On the other hand, LMIS allows the estimator bias to decrease with the number of outer samples. 
With larger $N$, more samples can initially be used to estimate the posterior moments used to find optimal biasing distributions. 
Furthermore, the LMIS estimator applies incremental enrichment a total of $N$ times, so as $N$ becomes larger, more samples are made available for use when estimating the posterior moments. 
When using LMIS, the variances of the inner estimators are significantly reduced, so the $1/NM$ terms in the variance for the LMIS estimator are negligible in comparison to prior-based sampling. 
Thus, instead of the usual bias--variance trade-off between maximizing the number of inner and outer samples, the earlier analysis and the numerical results suggest that maximizing the number of outer samples is the best way to reduce both the bias and variance!
In fact, if the biasing distributions converge to the posterior distributions as $N\to\infty$ (e.g., if they are of the same family of distributions), it is conceivable that one could set $M=1$ in this limit, since importance samping using the normalized posterior results in a zero-variance estimator. However, for finite sample sizes and non-Gaussian posterior distributions, we still require $M\to\infty$ to ensure asymptotic consistency, since the inner estimators correct for the mismatch between the biasing distributions and the posterior distributions.

\subsubsection{Importance sampling diagnostics and incremental enrichment} 
\begin{figure}[htbp]\centering
    \includegraphics[width=4.5in]{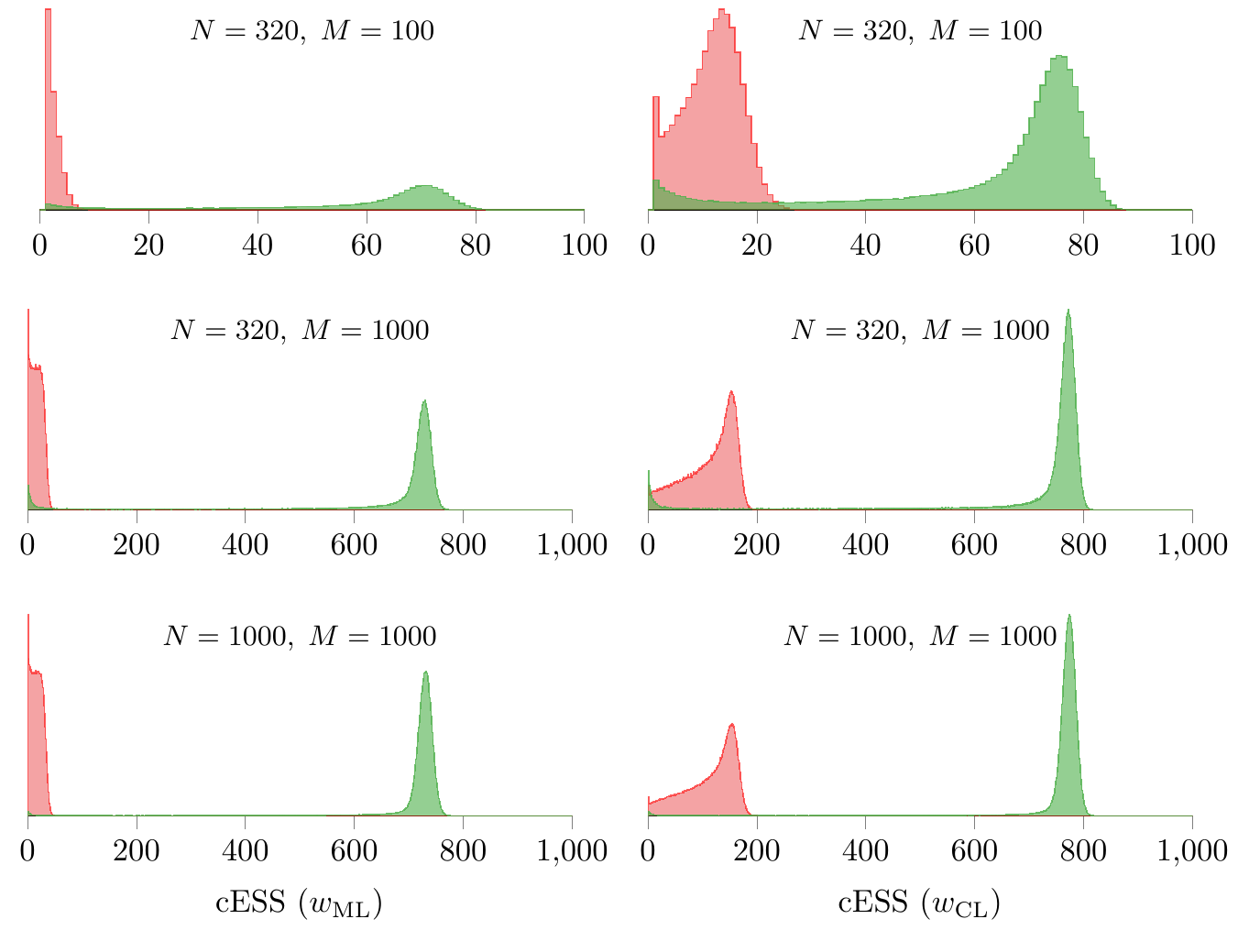}
    \caption{Histograms of `customized' effective sample sizes of the marginal and conditional likelihood estimators across 800 replicate simulations (Prior: red, LMIS: green). In each case, the effective sample size is bounded above by $M$.}%
    \label{fig:ess-hist}
\end{figure}

The exact value of the EIG is generally not available, so in practice, we must instead rely on importance sampling diagnostics to measure the performance of the estimator.
A useful ``rule of thumb'' for measuring importance sampling performance is to use the effective sample size (ESS) to measure how different the biasing distribution is from the target distribution \cite{liu2008monte}.
However, in our case, the target distribution is the prior distribution, so maximizing the ESS diagnostic would suggest choosing the prior as the biasing distribution, which we showed to be sub-optimal.
Instead, we consider the effective sample size customized for the target function---the marginal or conditional likelihood---as suggested in \cite{mcbook}. For a diagnostic that is specific to the importance sampling estimator in \cref{eq:marginal-likelihood-estimator}, we define
\begin{align*}
	\widetilde{w}_\text{ML}^{(i,j)}&\coloneqq\frac{p(\vy^{(i)}|\vtheta^{(i,j)},\veta^{(i,j)},\vd)\,p(\vtheta^{(i,j)},\veta^{(i,j)})/q_\text{ML}^{(i)}(\vtheta^{(i,j)},\veta^{(i,j)})}{\sum_{j=1}^{M_1}p(\vy|\vtheta^{(i,j)},\veta^{(i,j)},\vd)p(\vtheta^{(i,j)},\veta^{(i,j)})/q_\text{ML}^{(i)}(\vtheta^{(i,j)},\veta^{(i,j)})},\\
	\widetilde{w}_\text{CL}^{(i,j)}&\coloneqq\frac{p(\vy^{(i)}|\vtheta^{(i)},\veta^{(i,j)},\vd)\,p(\veta^{(i,j)}\given\vtheta^{(i)})/q_\text{CL}^{(i)}(\veta^{(i,j)})}{\sum_{j=1}^{M_1}p(\vy|\vtheta^{(i)},\veta^{(i,j)},\vd)p(\veta^{(i,j)}\given\vtheta^{(i)})/q_\text{CL}^{(i)}(\veta^{(i,j)})}.
\end{align*}
Then, the effective sample sizes customized to the marginal likelihood and the conditional likelihood are
\begin{align}
	\text{cESS}(w_\text{ML})=\frac{1}{\sum_{j=1}^{M_1}\bigl(\widetilde{w}_\text{ML}^{(i,j)}\bigr)^2},\quad
	\text{cESS}(w_\text{CL})=\frac{1}{\sum_{j=1}^{M_1}\bigl(\widetilde{w}_\text{CL}^{(i,j)}\bigr)^2}.
	\label{eq:cess}
\end{align}
In \cref{fig:ess-hist}, we can see that LMIS greatly increases the effective sample size customized to the marginal and conditional likelihood for both small $N=320$ and large $N=1000$ values of $N$ and $M$.

\begin{figure}[htbp]\centering
    \includegraphics[width=0.9\textwidth]{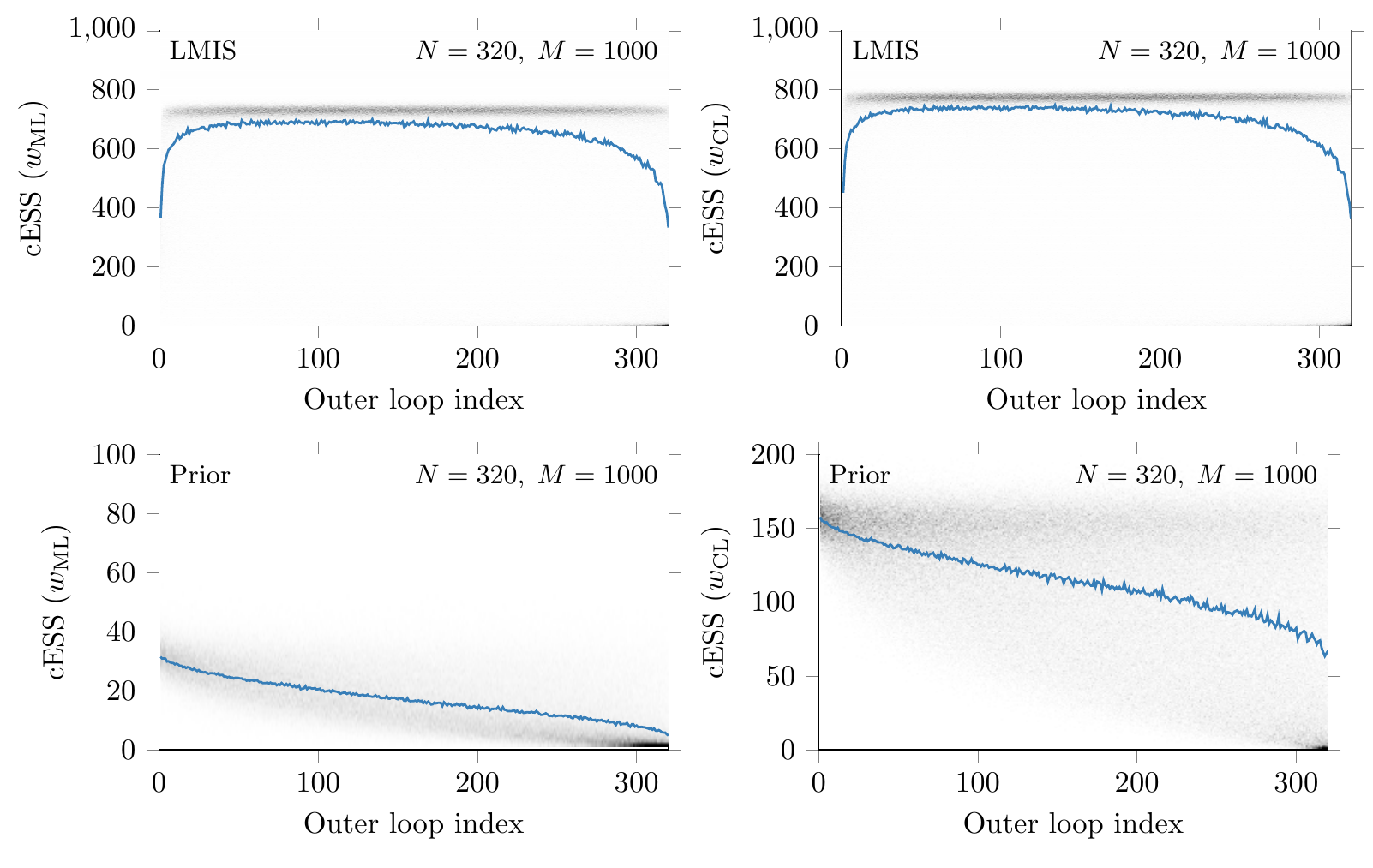}%
    \caption{A two-dimensional histogram of `customized' effective sample sizes of the marginal and conditional likelihood estimators, indexed by the outer loop iteration $i$, for 800 replicate simulations (i.e., there are 800 independent samples of the cESS for estimating conditional and marginal likelihood for each $\vy^{(i)}$). The solid line denotes the average cESS at each $i$. }
    \label{fig:ess-scatter}
\end{figure}

 In \cref{fig:ess-scatter}, we explore the mechanism by which incremental enrichment can improve the quality of the biasing distributions for the inner estimators. Since the samples of $(\vtheta^{(i)},\veta^{(i)})$ and the corresponding $\vy^{(i)}$ are sorted in decreasing order of prior density $p(\vtheta^{(i)},\veta^{(i)})$, we should expect that the effective sample size of the inner estimators decrease with increasing outer loop index $i$. This is particularly evident when using the prior as a biasing distribution, since as $i$ increases, the $i$th posterior $p(\vtheta,\veta\given\vy^{(i)},\vd)$ will be further out in the tail of the prior distribution $p(\vtheta,\veta)$, and is reflected in the monotonic decrease of cESS with $i$ seen in \cref{fig:ess-scatter}. On the other hand, when using LMIS, the cESS initially increases for small $i$, as a \emph{direct result of incremental enrichment.} Then, the cESS decays more slowly for the bulk of the inner estimates than in the prior case. Finally, for large $i$, the quality of the biasing distributions deteriorates rapidly, since there are fewer available samples for reuse when estimating the moments of posterior distributions concentrated in the tail of the prior. However, as $N$ becomes larger, the impact of the deterioration in the tails diminishes (cf.\ the decrease in probability mass in the lower-right corner of each subplot in \cref{fig:ess-scatter}). 
The deterioration in the tails is also reflected in \cref{fig:convergence_sweep} where the bias of the LMIS does not decrease as $1/M$ for fixed $N$. This can be attributed to the observation that the biasing distributions $q_\text{CL}^{(i)}$ for estimating the conditional likelihoods $p(\vy^{(i)}\given\vtheta,\vd)$ are obtained by calculating conditionals of an estimated posterior covariance. The conditional mean and variance are very sensitive to the correlation in the estimated posterior covariance, which may not be a robust estimator in the small sample size limit. The result is that the variance of the inner estimator of the conditional likelihood may decrease more slowly than $1/M$.

\subsection{Nonlinear example: M\"ossbauer spectroscopy}%
\label{ssec:mossbauer}
The M\"ossbauer effect refers to recoil-free nuclear resonance fluorescence, which involves the resonant and recoil-free emission and absorption of gamma radiation by atomic nuclei bound in a solid~\cite{mossbauer1958}. M\"ossbauer spectroscopy is responsible for the discovery of the M\"ossbauer isomeric shift~\cite{PhysRevLett.4.412}, which carries 
valuable information regarding the electronic and chemical structure of the lattice in which emitting and absorbing nuclei are embedded. These shifts are incredibly small, usually on the order of $10^{-8}$\,eV, which is roughly 12 orders of magnitude smaller than the energy of the emitted gamma rays themselves. Spectroscopic experiments involve moving the gamma ray source relative to the absorber. Change in the energy of the emitted gamma ray from Doppler shifting can be precisely controlled by the relative velocity of the source and absorber. In practice, shifts of $10^{-8}$\,eV correspond to velocities on the order of mm/s.

The general experimental setup for M\"ossbauer spectroscopy involves a radioisotope source, a drive system to move the source (either with constant acceleration or constant velocity), an absorber, and a gamma ray detector.
The M\"ossbauer drive is often operated at constant velocity for precision measurements, so that integration time may be spent only at the velocities of interest, which, in general, are not uniformly or symmetrically distributed about zero~\cite{mossbauerdrive}. If one uses long counting times and small velocity steps to obtain high precision, the total time of the run can take days, so careful selection of velocities at which to acquire data is critical for a successful experiment.
\begin{figure}[htbp]
  \subfloat[M\"ossbauer spectroscopy observation model]{%
    \includegraphics[width=0.58\linewidth]{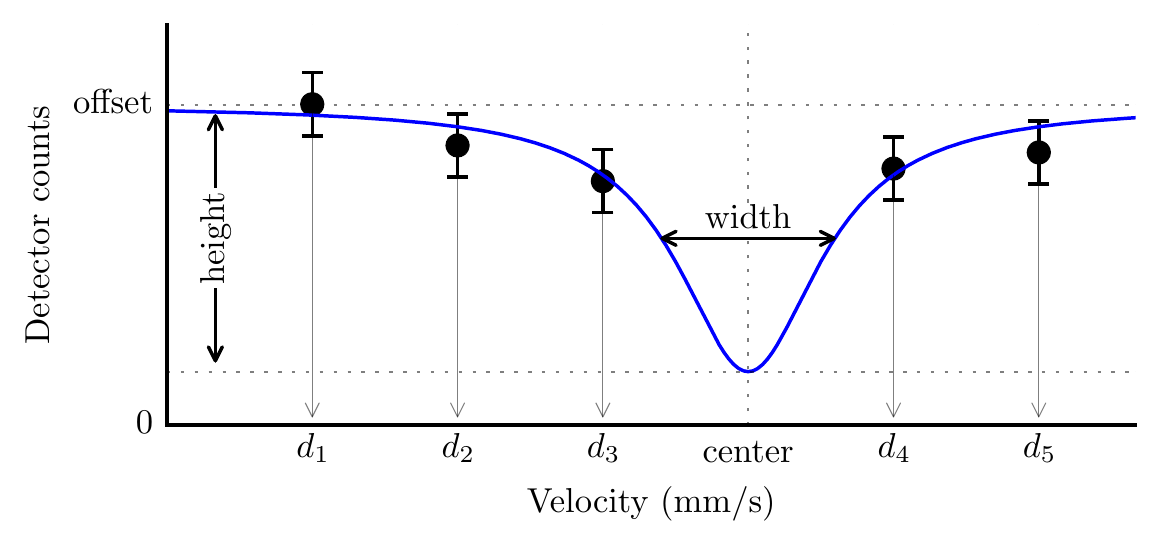}%
  }%
  \subfloat[Prior predictives ($n=50$)]{%
    \includegraphics[width=0.42\linewidth]{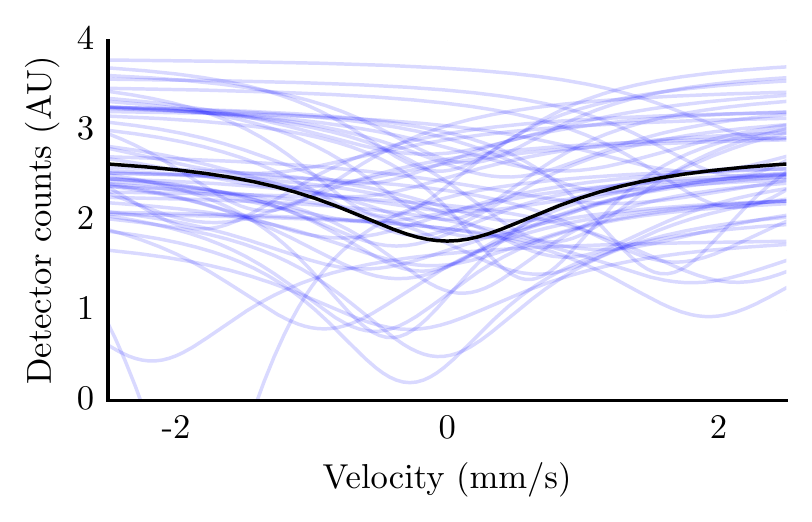}%
  }
  \caption{M\"ossbauer observation model and prior predictives}%
  \label{fig:mossbauer-obsmodel}
\end{figure}
\subsubsection{Problem formulation}
An experimental design consists of a set of $n_d$ velocities for high precision measurement of the isomeric shift---the center of the absorption peak---which we denote by $\theta$. 
We use the standard parameterization of the absorption peak as having a Lorentzian profile~\cite{mossbauersimple}, such that the the number of detector counts $y_i$ at a velocity $d_i$ can be described by the following relationship, also illustrated in \cref{fig:mossbauer-obsmodel}:
\begin{align}
y_i(d_i;\,\cdot \,) = \text{offset} - \text{height}\frac{\text{width}^2}{\text{width}^2 + {(\text{center}-d_i)}^2} +\epsilon_i,\quad i=1,2,\dots,n_d
\end{align}
The model parameters are assigned the following prior distributions,
\begin{eqnarray*}
  \text{center}\sim\mathcal{N}(0,1), &\quad
  \exp(\text{width})\sim\mathcal{N}(0,0.3^2),\\
  \exp(\text{height})\sim\mathcal{N}(0,0.3^2), &\quad
  \exp(\text{offset})\sim\mathcal{N}(1.0,0.2^2),
\end{eqnarray*}
where log-normal priors are assigned to parameters which can only have positive values, and the mean and variance of the prior distributions are chosen to ``standardize'' the signal, which would be standard experimental practice before collecting data. The additive observation error is assumed to be Gaussian  where $\epsilon_i\sim\mathcal{N}(0,0.1^2)$, which represents a signal-to-noise ratio of 10:1. 

\begin{figure}[h!]
    \subfloat[Expected utility for learning `center' with optimal designs marked by stars and illustrated above.]{%
      \begin{minipage}[t]{0.475\linewidth}\centering
      \includegraphics[width=0.95\linewidth]{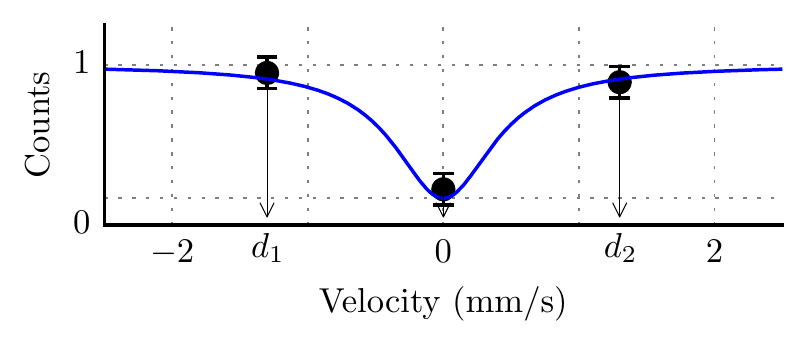}\\
      \includegraphics[width=0.95\linewidth]{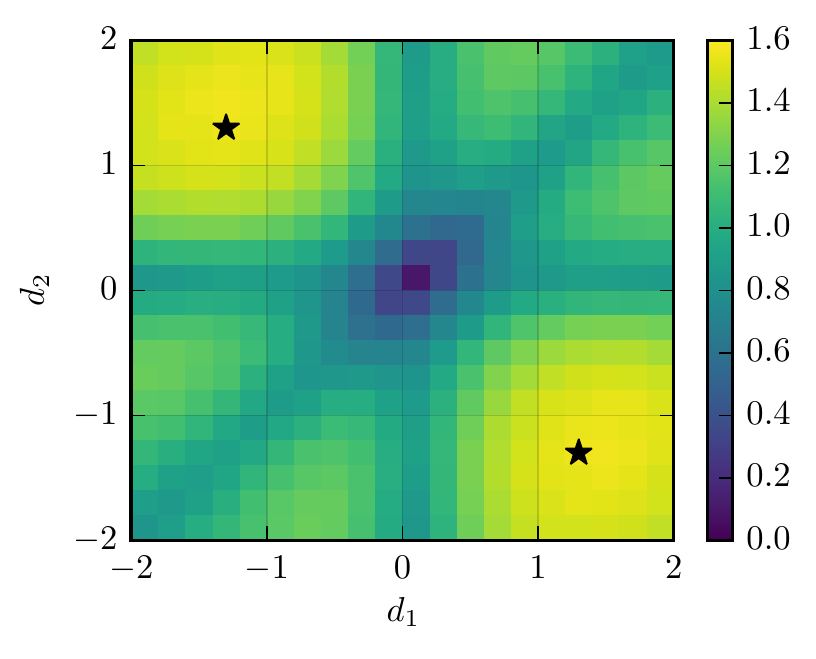}
      \end{minipage}
    }
    \subfloat[Expected utility for learning `offset' with optimal designs marked by stars and illustrated above.]{%
      \begin{minipage}[t]{0.475\linewidth}\centering
      \includegraphics[width=0.95\linewidth]{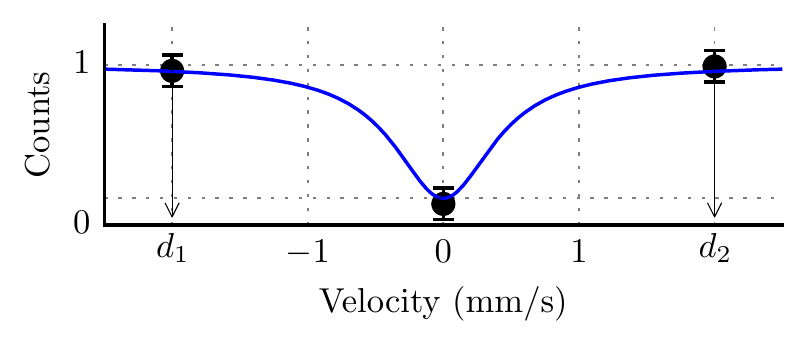}\\
      \includegraphics[width=0.95\linewidth]{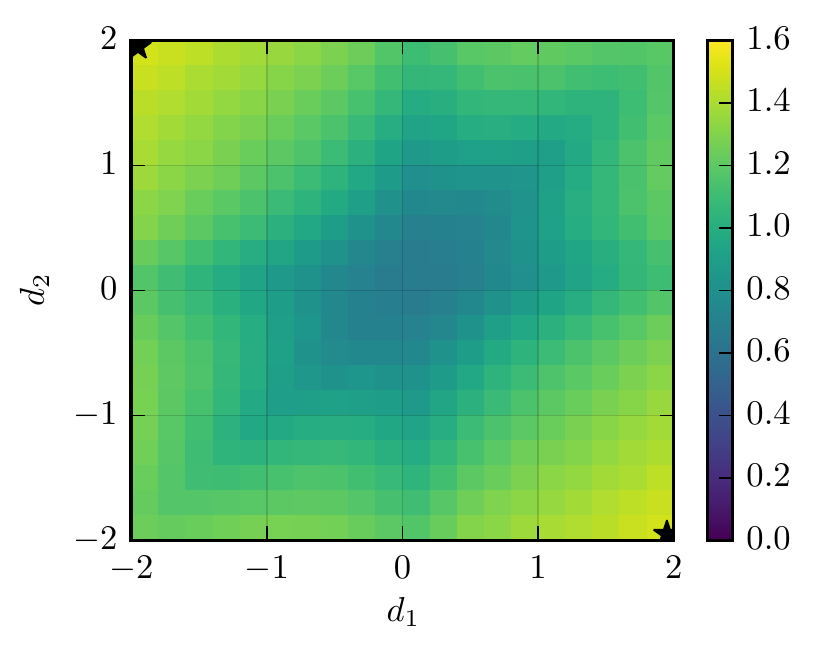}%
      \end{minipage} 
    }
    \caption{EIG for designs parameterized by $\vd=(0,d_1,d_2)$ estimated using LMIS with $N=1000$ and $M_1=M_2=100$ samples.}%
  \label{fig:mossbauer-utility}
\end{figure}
\begin{figure}[htbp]
    \centering
    \includegraphics[width=5in]{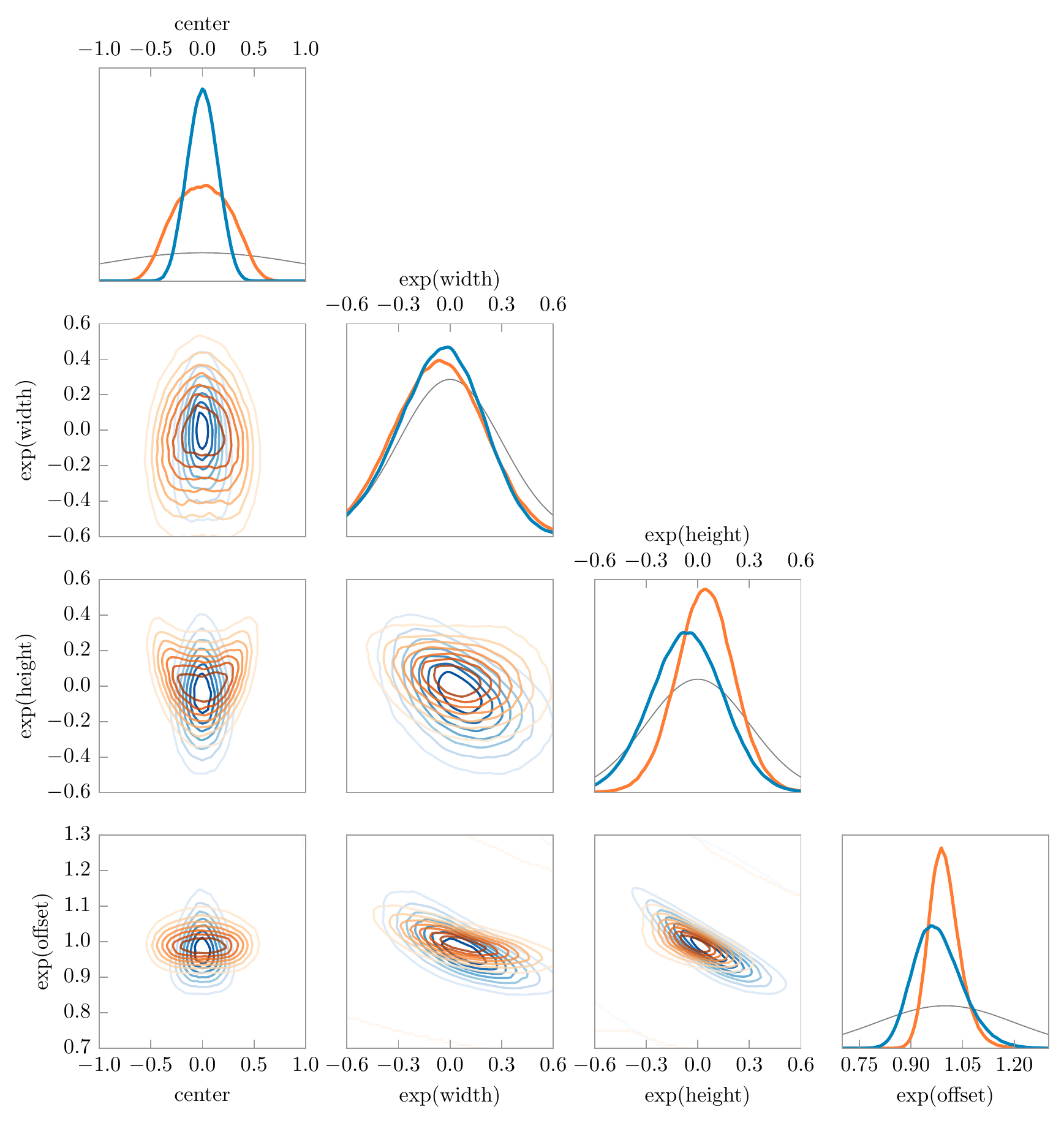}
    \caption{Marginals of the 4-dimensional posterior distributions resulting from two experimental designs with different design objectives. Blue: optimal design for `center' parameter: $\boldsymbol{d} = (-1.3,0,1.3)$. Orange: optimal design for `offset' parameter $\boldsymbol{d} = (-2,0,2)$. The marginal prior densities are drawn in gray.}%
    \label{fig:mossbauer-overlay}
\end{figure}

In \cref{fig:mossbauer-utility}, we plot two ``maps'' of EIG 
estimated using LMIS, using sample sizes $M_1 = M_2 = 100$ and $N= 1000$. 
One map uses $\theta = \text{`center'}$ and lets $\eta$ comprise the remaining parameters; the other map puts $\theta = \text{`offset'}$ and lets $\eta$ comprise the rest. The optimal designs differ for these two scenarios: $\boldsymbol{d}_\text{center}^\ast = (-1.3,0,1.3)$ and $\boldsymbol{d}_\text{offset}^\ast = (-2,0,2)$. For visualization purposes, we restrict the design problem to three velocities, with one of the three velocities fixed to zero. To better understand the impact of these design choices, \Cref{fig:mossbauer-overlay} plots the posterior distribution obtained using a data realization from each optimal design. Comparing the two focused designs, we observe a distinct tradeoff between learning about the `center' parameter and the `offset' parameter, especially in the lower-left pairwise marginal. Furthermore, the two-dimensional marginals clearly illustrate the non-Gaussianity of the posterior distribution.

\begin{figure}[h!]
    \centering
    \includegraphics[width=5.5in]{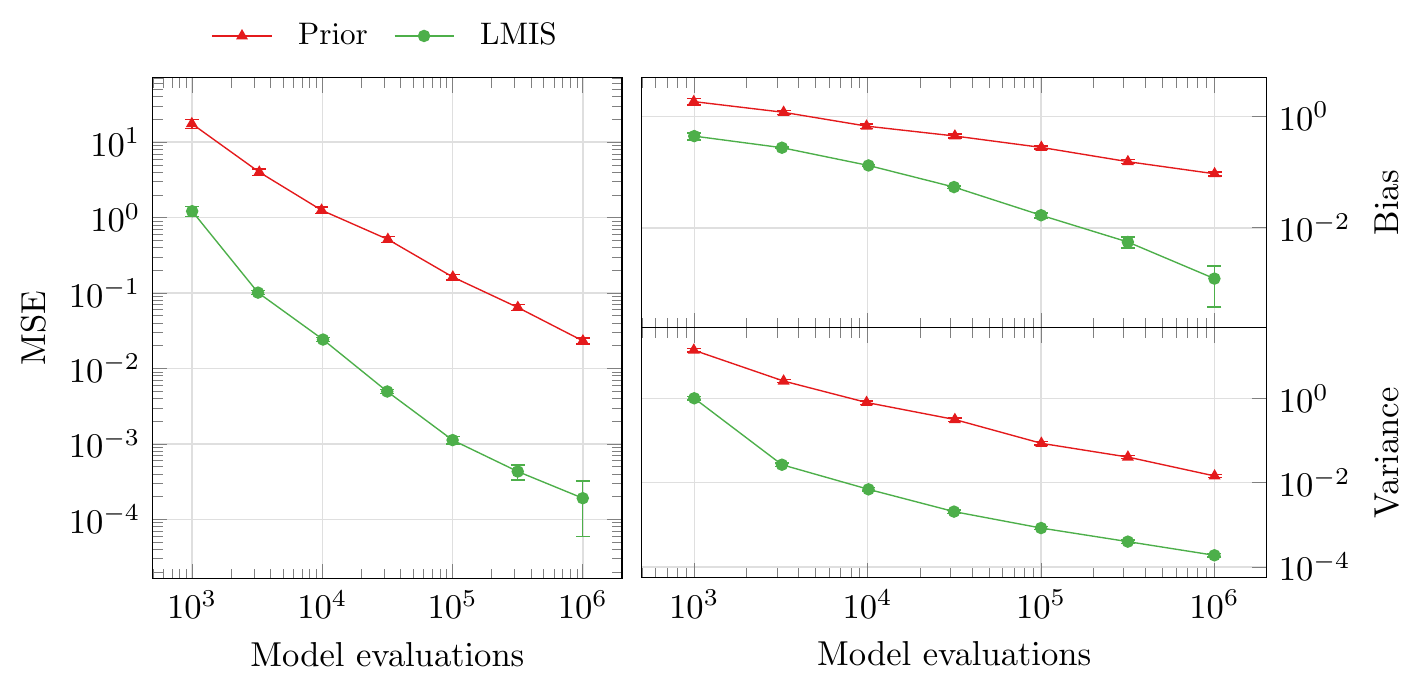} %
    \caption{M\"o{}ssbauer Example. (Left) Mean squared error of the estimate of the EIG in the `center' parameter for the design $\vd_\text{center}^\ast=(-1.3, 0, 1.3)$. Error bars denote the Monte Carlo standard error from averaging over 100 replicates.}%
    \label{fig:mossbauer-convergence}
\end{figure}

Next we evaluate the performance of the LMIS estimator, focusing on the estimation of EIG in the marginal distribution of $\theta = \text{`center.'}$ Since we do not have a closed-form expression for this EIG, we resort to calculating a ``reference'' value by using a large computational budget: $10^8$ model evaluations with the LMIS estimator. In \cref{fig:mossbauer-convergence}, we plot the mean squared error with respect to this reference value for both the LMIS estimator and the prior estimator. We choose to evaluate the focused EIG for the design  $\vd_\text{center}^\ast=(-1.3, 0, 1.3)$ since it presents the most challenging case---with the greatest amount of posterior concentration. Compared to the prior estimator, we observe a reduction of up to two orders of magnitude in the mean squared error when using LMIS. This result is consistent with the earlier linear-Gaussian example and suggests that the algorithm is able to identify high-quality biasing distributions even in the nonlinear case.

\begin{figure}[htbp]\centering
  \includegraphics[width=5in]{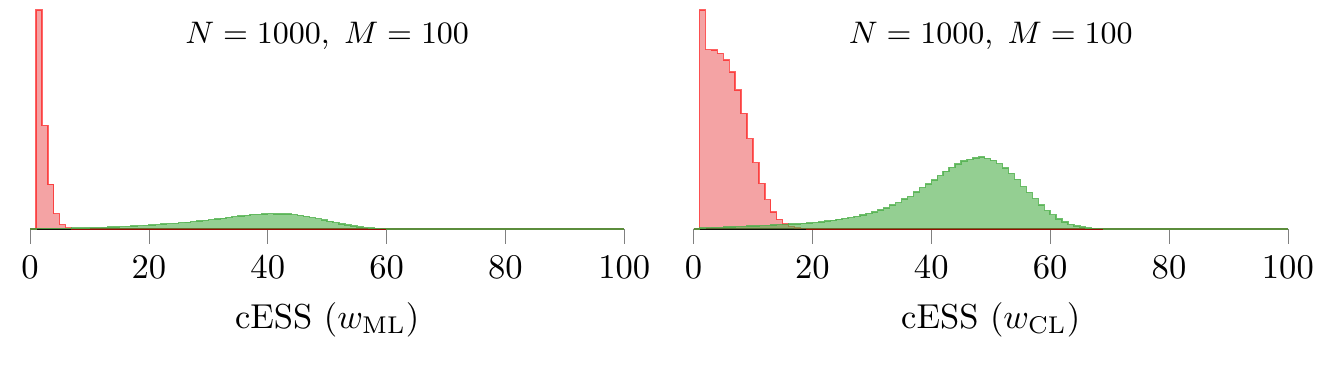}
  \caption{M\"o{}ssbauer example with design $\boldsymbol{d}=(-1.3,0,1.3)$. Histograms of `customized' effective sample sizes of the marginal and conditional likelihood estimators across 800 replicate simulations (Prior: red, LMIS: green). In each case, the effective sample size is bounded above by $M$. }%
  \label{fig:mossbauer-ess}
\end{figure}

We can also use the cESS diagnostic \cref{eq:cess} to evaluate the performance of the LMIS scheme. The distribution of cESS is plotted in \cref{fig:mossbauer-ess}, and displays a marked increase over that obtained when using the prior as a biasing distribution, comparable to the improvement for the linear Gaussian example of the same dimension. %
Again, this result suggests that LMIS is able to find high-quality biasing distributions for estimating both the marginal and conditional likelihoods, even in the case of a nonlinear observational model and non-Gaussian posteriors.

\section{Conclusion}%
\label{sec:conclusion}
This paper introduces a new computational approach for \textit{focused} optimal Bayesian experimental design, where the design objective is to maximize expected information gain in the marginal distribution of parameters of interest. In particular, we develop a consistent layered multiple importance sampling (LMIS) estimator of this objective, which uses incremental enrichment to construct a sequence of increasingly tailored biasing distributions.
We demonstrate the effectiveness of our approach on design problems of moderate dimension. Compared to previous algorithms that draw samples from the prior, the LMIS estimator reduces both bias and variance by several orders of magnitude, for the same number of model evaluations. The reduction in variance may improve the efficiency of optimization over the design space, but more importantly, the reduction in bias diminishes the confounding effect caused by link between estimator bias and posterior concentration. The latter is particularly important in the context of focused design, as posterior concentration in general has a different dependence on the design parameters than the information gain in selected \textit{marginals} of interest.

The enrichment process of LMIS suggests a different balance of inner-loop and outer-loop samples than previous nested Monte Carlo methods. We also emphasize that LMIS does not require gradients of the likelihood or deterministic models embedded therein, and thus is well suited to problems where gradient information may not be available. While here we implemented LMIS with a particular family of inner-loop biasing distributions (multivariate $t$), the overall algorithm can be used with any parametric family or mixture, as suited to the form of the posterior.

Possible extensions of this work include using surrogate models to accelerate estimation; while previous work did so in a biased way \cite{huan2013}, the LMIS framework could incorporate surrogate models only in its outer layer, to help construct biasing distributions more cheaply. When gradients of the posterior density are readily available, one also could use Laplace approximations of the posterior distribution to construct suitable biasing distributions for estimation of the marginal \cite{Beck2017} and conditional likelihoods. Moreover, surrogate models and approximations of the marginal likelihood could be combined with the full-model LMIS estimator using a control variate or multilevel Monte Carlo approach, to further reduce the number of full model evaluations while preserving consistency. Another clear extension is to pair LMIS estimators with optimization methods designed for stochastic objectives (e.g.,~\cite{XunIJUQ, WeaverBayesianAnalysis, OverstallWoodsApproximateCoordinateExchange,carlon2018nesterov}) to facilitate search over the design space. 

An interesting application of focused experimental design is for models that contain discrepancy terms \cite{kennedy:2001:bco} intended to capture model misspecification. In this setting, one could treat parameters of the discrepancy model as nuisance parameters ($\eta$), where the goal is to learn about other parameters of interest ($\theta$) in the presence of model discrepancy. Alternatively, one could focus on learning about the discrepancy term to better capture the nature of the mismatch between the model and data.

\appendix

\section{Asymptotic bias and variance of the EIG estimator}\label{sec:appendix}
In this section, we determine the asymptotic behavior of the bias and
variance of the expected information gain (EIG) estimator $\widehat{U}(\vd)$ \eqref{eq:big-mc-estimator} with respect to the number of samples in both the inner and outer sums.
Let $A:\mathbb{R}^{n_\eta}\to\mathbb{R}^+$ and $B:\mathbb{R}^{n_\eta}\times\mathbb{R}^{n_\theta}\to\mathbb{R}^+$ be scalar functions defined as
\begin{align*}
  A(\veta;\underline{\vy},\underline{\vtheta},\underline{\vd})&\coloneqq p(\underline{\vy}\given\underline{\vtheta},\veta,\underline{\vd}),\\
  B(\vtheta,\veta;\underline{\vy},\underline{\vd})&\coloneqq p(\underline{\vy}\given{\vtheta},\veta,\underline{\vd}),
\end{align*}
where underlined variables are fixed parameters. 
The expectations of these functions are equal to the conditional and
marginal likelihoods, respectively:
\begin{align*}
\begin{array}{rclcl}
\mu_{A}(\vy,\vtheta;\vd)&\coloneqq& \Ex_{\veta \vert \vtheta}[A(\veta;\vy,\vtheta,\vd)]&=&p(\vy\given\vtheta,\vd),\\
\mu_{B}(\vy;\vd)&\coloneqq& \Ex_{\vtheta,\veta}[B(\vtheta,\veta;\vy,\vd)]&=&p(\vy\given\vd).
\end{array}
\end{align*}
Let $\SA={\{\veta^{(k)}\}}_{k=1}^{M_2}$ and $\SB={\{(\vtheta^{(j)},\veta^{(j)})\}}_{j=1}^{M_1}$ be samples drawn i.i.d.\ from the biasing distributions $q_\text{cond}^{(\vy,\vtheta)}(\veta)$ and $q_\text{marg}^{(\vy)}(\vtheta,\veta)$, respectively.
Then, the importance sampling estimates of the conditional and marginal likelihoods from \cref{eq:marginal-likelihood-estimator,eq:conditional-likelihood-estimator} can be written as
\begin{align*}
\widehat{A}(\vy,\vtheta,\SA;\vd) &= \frac{1}{M_2}\sum_{j=1}^{M_2} A(\veta^{(j)};\vy,\vtheta,\vd)\frac{p(\veta^{(k)}\given\vtheta^{(k)})}{q_\text{cond}^{(\vy,\vtheta)}(\veta^{(k)})},\\
\widehat{B}(\vy,\SB;\vd) &= \frac{1}{M_1}\sum_{j=1}^{M_1} B(\vtheta^{(j)},\veta^{(j)};\vy,\vd)\frac{p(\vtheta^{(j)},\veta^{(j)})}{q_\text{marg}^{(\vy)}(\vtheta^{(j)},\veta^{(j)})},
\end{align*}
Now, define
\begin{align*}
u(\vy,\vtheta,\SA,\SB;\vd) \coloneqq \ln\frac{\widehat{A}(\vy,\vtheta,\SA;\vd)}{\widehat{B}(\vy,\SB;\vd)}
\end{align*}
so that $\widehat{U}(\vd)$ from~\eqref{eq:marginal-likelihood-estimator} can be written as
\begin{align*}
\widehat{U}(\vd) = \frac{1}{N}\sum_{i=1}^N u(\vy^{(i)},\vtheta^{(i)},S_A^{(i)},S_B^{(i)};\vd)
\end{align*}
where $\vy^{(i)}\sim p(\vy^{(i)}\given\vtheta^{(i)},\veta^{(i)},\vd)$, $(\vtheta^{(i)},\veta^{(i)})\sim p(\vtheta,\veta)$,
and $S_A^{(i)}$ and $S_B^{(i)}$ are sets of i.i.d.\ samples drawn from $q_\text{cond}^{(\vy^{(i)},\vtheta^{(i)})}$ and $q_\text{marg}^{(\vy^{(i)})}$, respectively.

From the law of iterated expectations, the expected value of $\widehat{U}(\vd)$ can be written as
\begin{align}
  \Ex[\widehat{U}(\vd)]&=\Ex_{\vy,\vtheta}\bigl[\Ex_{\SA,\SB}[u(\vy,\vtheta,\SA,\SB;\vd)]\bigr],
  \label{eq:expected-value-Uhat}
\intertext{and from the law of total variance, the variance of $\widehat{U}(\vd)$ can be written as}
  \Var[\widehat{U}(\vd)]&=\frac{1}{N}\Var_{\vy,\vtheta,\SA,\SB}[u(\vtheta,\vy,\SA,\SB)]\nonumber \\
  &=\frac{1}{N}\Ex_{\vy,\vtheta}\bigl[\Var_{\SA,\SB}[u(\vy,\vtheta,\SA,\SB;\vd)]\bigr] + \frac{1}{N}\Var_{\vy,\vtheta}\bigl[\Ex_{\SA,\SB}[u(\vy,\vtheta,\SA,\SB;\vd)]\bigr].
  \label{eq:variance-Uhat}
\end{align}
Using the delta method, we can obtain the second-order approximation of the expected value and variance of $u(\vy,\vtheta,\SA,\SB;\vd)$:
\begin{align}
\label{eq:expected-value-delta}
\Ex_{\SA,\SB}[u(\vy,\vtheta,\SA,\SB;\vd)]
&=\ln\frac{\mu_{A}(\vy,\vtheta)}{\mu_{B}(\vy)}+\frac{1}{2}\biggl[\frac{1}{M_1}\frac{\sigma_B^2(\vy;\vd)}{\mu_{B}^2(\vy;\vd)}-\frac{1}{M_2}\frac{\sigma_A^2(\vy,\vtheta;\vd)}{\mu_{A}^2(\vy,\vtheta;\vd)}\biggr]\\\nonumber
&\qquad+\mathcal{O}(1/M_1^2) + \mathcal{O}(1/M_2^2),\\
\label{eq:variance-delta}
\Var_{\SA,\SB}[u(\vy,\vtheta,\SA,\SB;\vd)]
&=\frac{1}{M_2}\frac{\sigma_A^4(\vy,\vtheta;\vd)}{\mu_{A}^4(\vy;\vd)}+\frac{1}{M_1}\frac{\sigma_B^4(\vy,\vtheta;\vd)}{\mu_{B}^4(\vy;\vd)}\\\nonumber
&\qquad+\mathcal{O}(1/M_1^2) + \mathcal{O}(1/M_2^2),
\end{align}
where we have used the fact that $\widehat{A}(\vy,\vtheta;\vd)$ and $\widehat{B}(\vy;\vd)$ are unbiased estimators, so their expected values are
\begin{align*}
\Ex_{\SA,\SB}[\widehat{A}(\vy,\vtheta,\SA;\vd)] &= \mu_{A}(\vy,\vtheta;\vd), \\
\Ex_{\SA,\SB}[\widehat{B}(\vy,\SB;\vd)]&=\mu_{B}(\vy;\vd),
\end{align*}
and their variances are
\begin{align*}
\Var_{\SA,\SB}[\widehat{A}(\vy,\vtheta,\SA;\vd)] &=
\frac{1}{M_2}
\Var_{{\veta}\sim q_\text{cond}^{(\vy, \vtheta)}}\biggl[A(\veta;\vy,\vtheta,\vd)\frac{p({\veta}\given\vtheta)}{q_\text{cond}^{(\vy, \vtheta)}({\veta})}\biggr]
=\frac{\sigma_{A}^2(\vy,\vtheta;\vd)}{M_2},\\
\Var_{\SA,\SB}[\widehat{B}(\vy,\SB;\vd)] &=
\frac{1}{M_1}
\Var_{({\vtheta},{\veta})\sim q_\text{marg}^{(\vy)}}\biggl[B(\vtheta,\veta;\vy,\vd)\frac{p({\vtheta},{\veta})}{q_\text{marg}^{(\vy)}({\vtheta},{\veta})}\biggr]
=\frac{\sigma_{B}^2(\vy;\vd)}{M_1}.
\end{align*}
Then, the expected value of $\widehat{U}(\vd)$ from~\eqref{eq:expected-value-Uhat} becomes
\begin{align}
  \Ex[\widehat{U}(\vd)] &=
  \underbrace{\Ex_{\vy,\vtheta}\biggl[
    \ln\frac{\mu_{A}(\vy,\vtheta;\vd)}{\mu_{B}(\vy;\vd)}
  \biggr]}_{U(\vd)}
   + \frac{1}{M_1}\underbrace{\Ex_{\vy}\biggl[
    \frac{\sigma_B^2(\vY;\vd)}{2\mu_{B}^2(\vY;\vd)}
   \biggr]}_{C_1(\vd)}
   - \frac{1}{M_2}\underbrace{\Ex_{\vy,\vtheta}\biggl[
    \frac{\sigma_A^2(\vy,\vtheta;\vd)}{2\mu_{A}^2(\vy,\vtheta;\vd)} 
   \biggr]}_{C_2(\vd)}\\
   &\qquad+\mathcal{O}(1/M_1^2) + \mathcal{O}(1/M_2^2)\nonumber \\
   &= U(\vd) + \frac{C_1(\vd)}{M_1}-\frac{C_2(\vd)}{M_2}+\mathcal{O}(1/M_1^2) + \mathcal{O}(1/M_2^2),
\end{align}
And thus the bias of $\widehat{U}(\vd)$ is
\begin{align}
  \Ex[\widehat{U}(\vd)-U(\vd)] = \frac{C_1(\vd)}{M_1} - \frac{C_2(\vd)}{M_2}+\mathcal{O}(1/M_1^2) + \mathcal{O}(1/M_2^2).
  \label{eq:Uhat-bias}
\end{align}
Similarly, the variance of $\widehat{U}(\vd)$ from~\eqref{eq:variance-Uhat} becomes
\begin{align}
  \Var[\widehat{U}(\vd)] &\approx \frac{1}{N}\biggl(\frac{1}{M_2}\overbrace{\Ex_{\vy,\vtheta}\biggl[\frac{\sigma_A^4(\vy,\vtheta;\vd)}{\mu_A^4(\vy,\vtheta;\vd)}\biggr]}^{D_1(\vd)}+\frac{1}{M_1}\overbrace{\Ex_{\vy}\biggl[\frac{\sigma_B^4(\vy;\vd)}{\mu_B^4(\vy;\vd)}\biggr]}^{D_2(\vd)}+\mathcal{O}(1/M_1^2) + \mathcal{O}(1/M_2^2)\biggr)+\nonumber \\
  &\quad\quad \frac{1}{N}\biggl(\underbrace{\Var_{\vy,\vtheta}\biggl[\ln\frac{\mu_A(\vy,\vtheta;\vd)}{\mu_B(\vy;\vd)}\biggr]}_{D_3(\vd)} +\frac{1}{{(M_1)}^2}\underbrace{\Var_{\vy}\biggl[\frac{\sigma_B^2(\vy;\vd)}{2\mu_B^2(\vy;\vd)}\biggr]}_{D_4(\vd)}+\frac{1}{{(M_2)}^2}\underbrace{\Var_{\vy,\vtheta}\biggl[\frac{\sigma_A^2(\vy,\vtheta;\vd)}{2\mu_A^2(\vy,\vtheta;\vd)}\biggr]}_{D_5(\vd)}\\
  &\qquad\qquad\qquad +\mathcal{O}(1/M_1^4) + \mathcal{O}(1/M_2^4)\biggr)\nonumber \\
  &=\frac{D_3(\vd)}{N} + \frac{D_1(\vd)}{NM_2}  + \frac{D_2(\vd)}{NM_1}  + \frac{D_4(\vd)}{N{(M_2)}^2}  + \frac{D_5(\vd)}{N{(M_1)}^2} + \mathcal{O}\bigl(\tfrac{1}{NM_1^3}\bigr)+\mathcal{O}\biggl(\frac{1}{NM_2^3}\biggr).
  \label{eq:Uhat-variance}
\end{align}

\section{Sample sizes and scaling}\label{appendix:scaling}
The nested estimator \eqref{eq:big-mc-estimator} is consistent (see \cite{rainforth2016nestedMC}) and in particular, its bias and variance converge to zero as $M_1, M_2,  N \to \infty$. Since all three sample sizes must increase, it is natural to ask whether there is an optimal allocation of inner ($M_1, M_2$) and outer ($N)$ samples for a given computational budget $W = N(M_1 + M_2)$.

For simplicity, let $M_1 = M_2 = M$ and define  $\widetilde{C} \coloneqq (C_1- C_2)^2>0$ and $\widetilde{D} \coloneqq D_1+D_2$, where we drop dependence on $\vd$ to reduce clutter. Then $W = 2MN$ and the bias and variance from \eqref{eq:delta-bias} and \eqref{eq:delta-variance} can be written as
\begin{align*}
    \text{bias} & \sim \frac{{\widetilde{C}}^{1/2}}{M},\qquad
    \text{variance} \sim \frac{D_3}{N}+\frac{\widetilde{D}}{NM},
\end{align*}
yielding an expression for the mean squared error (MSE) of the estimator, to leading order, in terms of $M$ and $N$:
\begin{align*}
    \text{MSE} & = \mathbb{E} \left [ \vert \widehat{U} - U \vert^2 \right ] = \text{bias}^2 + \text{variance} \sim \frac{\widetilde{C}}{M^2}+\frac{D_3}{N}+\frac{\widetilde{D}}{MN}.
\end{align*}
Let $\alpha^2=2M/N$ be the ratio between the number of inner and outer Monte Carlo samples. Then, by substituting $M=N\alpha^2/2$ and $N=\sqrt{W}/\alpha$, %
we can rewrite the MSE in terms of the total computational work $W$ and $\alpha$:
\begin{align}
    \text{MSE}(W,\alpha) & \sim \frac{4\widetilde{C}}{\alpha^2 W}+\frac{\alpha D_3}{W^{1/2}}+\frac{2\widetilde{D}}{W}.
    \label{eq:work-mse}
\end{align}
The value of $\alpha$ and the corresponding ratio $\alpha^2$ that minimize the leading-order MSE are
\begin{align}
    \alpha_\ast=2 \left ( \frac{\widetilde{C}}{D_3} \right )^{1/3} \frac{1}{W^{1/6}}
    \quad
    \Rightarrow\quad
    \alpha_\ast^2= 4 \left ( \frac{\widetilde{C}}{D_3} \right )^{2/3} \frac{1}{W^{1/3}}.
    \label{eq:work-alpha-opt-appendix}
\end{align}
This result suggests that the ratio between the number of inner and outer samples should decrease slowly as the computational budget increases, in particular scaling as $W^{-1/3}$. The ratio $\widetilde{C}/D_3$ reflects the relative magnitudes of leading-order terms in the bias and variance; if the bias dominates, one should choose a relatively larger number of inner loop samples. The minimum MSE, realized by this value of $\alpha_{\ast}$, is
\begin{align*}
    \text{MSE}(W,\alpha_\ast) \sim \frac{2\widetilde{D}}{W}+\frac{3\widetilde{C}^{1/3}D_3^{2/3}}{W^{2/3}}.
\end{align*}
Of course, if one does not know the constant in \eqref{eq:work-alpha-opt-appendix} and instead just uses $\alpha = \alpha_0/W^{1/6}$, the work-scaling of the MSE remains the same: 
\begin{align*}
    \text{MSE}(W, \alpha_0/W^{1/6} ) \sim \frac{2\widetilde{D}}{W} + \left ( \frac{4 \widetilde{C}}{\alpha_0^2} + \alpha_0 D_3 \right ) \frac{1}{W^{2/3}}
\end{align*}
Thus we can compare two scenarios. For a fixed ratio between $M$ and $N$, i.e., if  $\alpha^2$ were not allowed to depend on $W$, the leading-order term of \eqref{eq:work-mse} would yield $\text{MSE} = \mathcal{O}(W^{-1/2})$. If, instead, we adjust the ratio according to the optimal scaling $\alpha^2 \propto W^{-1/3}$, we obtain a faster decay rate of $\text{MSE} = \mathcal{O}(W^{-2/3})$.

\begin{figure}[htbp]
  \includegraphics[width=0.5\textwidth]{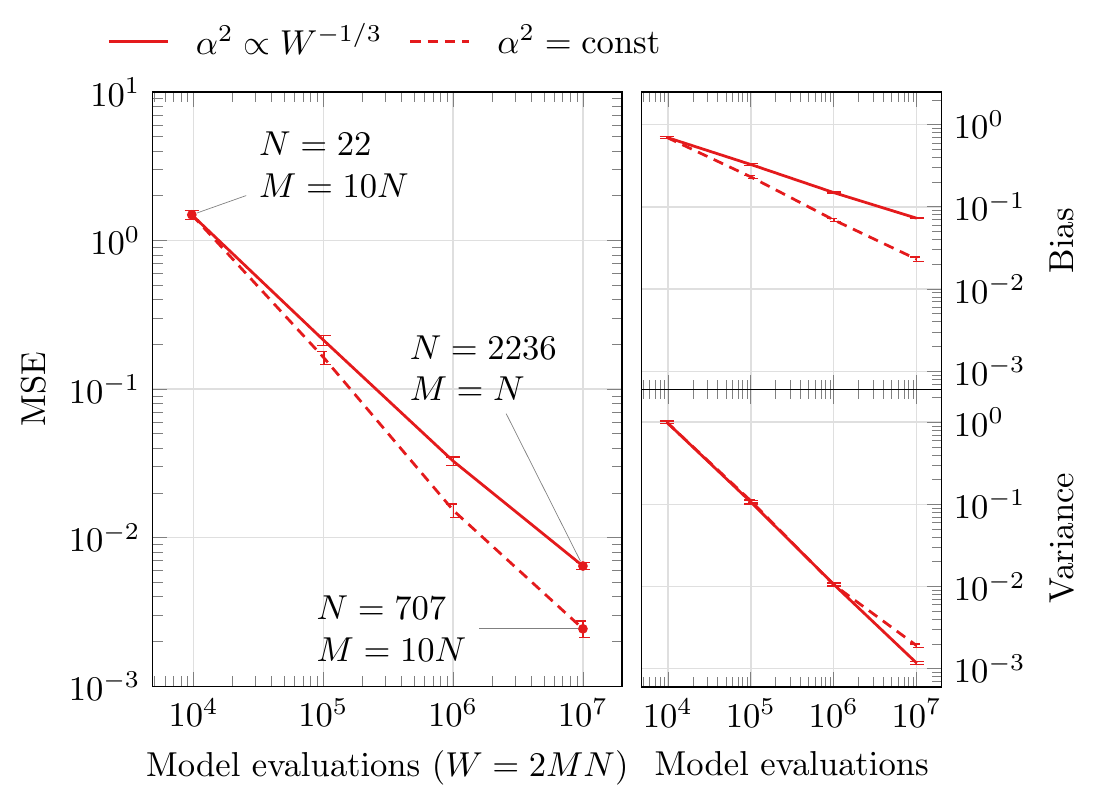}%
  \includegraphics[width=0.5\textwidth]{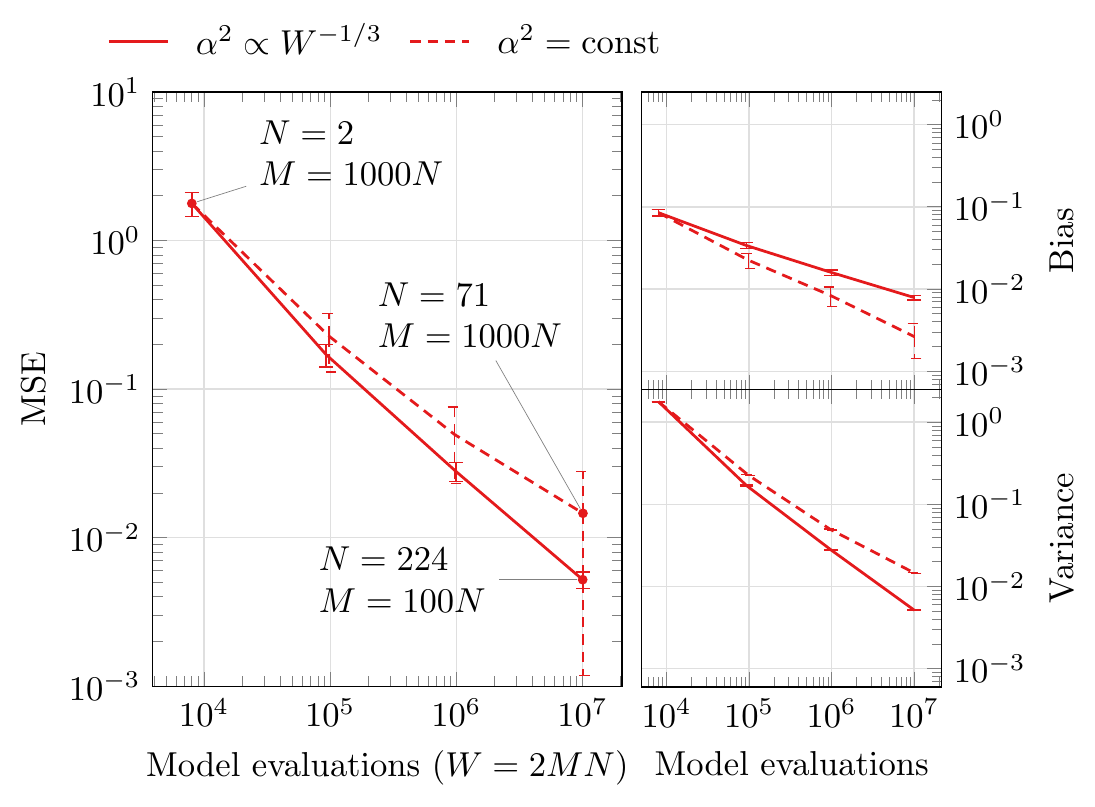}
  \caption{Mean square error of the expected information gain estimator as a function of model evaluations with different scaling strategies: optimal scaling (solid line) and fixed ratio (dashed line)}
  \label{fig:scaling-mse}
\end{figure}

We use the 4-dimensional linear Gaussian problem to verify the results of the optimal scaling analysis in \cref{ssec:biasvariancescaling}.
In \cref{fig:scaling-mse}, we plot the estimator MSE against the number of model evaluations using two scaling strategies: a fixed ratio (dashed line) and a scaled ratio (solid line).
In practice, the constant term $(\frac{\widetilde{C}}{D_3})^{2/3}$ in \cref{eq:work-alpha-opt-appendix} for the optimal ratio is not known.
In \cref{fig:scaling-mse}, we anchor the ratio at $W=10^4$ to be $M=10N$ (left) and $M=1000N$ (right) to investigate the influence of the constant term on the convergence rate for finite sample sizes.
We observe that when the MSE is dominated by the estimator bias (e.g., $M=10N$), the trade-off explored by the optimal scaling actually results in sub-optimal convergence in MSE.
This is because the benefits of a faster convergence rate in the estimator variance are overshadowed by the slower convergence rate in the bias.
On the other hand, when the bias and variance are of similar magnitude (e.g., $M=1000N$ at $W=10^4$), we can observe benefits of a faster convergence rate.
Between $W=10^6$ and $W=10^7$, the MSE converges at the predicted rates: $W^{-1/2}$ when using a fixed ratio and $W^{-2/3}$ when using the optimal scaling.
In practice, without knowing the optimal value of the constant from \cref{eq:work-alpha-opt-appendix} or the magnitude of the estimator bias, a strategy that uses a fixed ratio between the number of samples in the inner and outer estimators may be appropriate for realistic computational budgets.

\bibliographystyle{siamplain}
\bibliography{main}
\end{document}